\documentclass[manuscript,screen,nonacm]{acmart}\usepackage{longtable}
\usepackage{tcolorbox}
\tcbuselibrary{breakable}
\usepackage{tabularx}
\AtBeginDocument{%
  }

\setcopyright{none}
\renewcommand\footnotetextcopyrightpermission[1]{}

\usepackage{xcolor}
\usepackage{placeins}

\usepackage[disable]{todonotes}
\usetikzlibrary{arrows.meta}

\begin{document}

\title{Necessary for What? Auditing Demographic Disclosure to LLM Synthetic Respondents}

\author{Aspen Wang}
\affiliation{%
  \institution{Institute of Information Science, Academia Sinica}
  \city{Taipei}
  \country{Taiwan}}
\email{aspen@as.edu.tw}

\author{Yen Yu Shan}
\affiliation{%
  \institution{Institute of Information Science, Academia Sinica}
  \city{Taipei}
  \country{Taiwan}}
  \email{bubu3013@as.edu.tw}

\author{Lun-Wei Ku}
\affiliation{%
  \institution{Institute of Information Science, Academia Sinica}
  \city{Taipei}
  \country{Taiwan}}
    \email{lwku@iis.sinica.edu.tw}

\begin{abstract}
Large language models (LLMs) increasingly answer surveys on behalf of real panelists, and each panelist’s recorded demographic attributes (e.g., age, education, ethnicity) are often placed in the prompt on the premise that richer profiles make more faithful synthetic respondents. While data minimization limits personal data to what a purpose requires, a synthetic survey serves several purposes at once, from predicting individual answers to estimating how opinions are distributed and comparing groups. We ask which of these purposes require the respondent’s record, and whether improved estimates support sound conclusions about groups. We vary whose profile a model receives, where it comes from, and how many fields it contains, in two U.S. surveys with eleven open-weight and three hosted models, and compare each model with a statistical model fit to the same information. We make four contributions. First, we show that the record adds little to individual predictions and that, for the open-weight models, most of its gain in population estimates remains when it is replaced by census data or a synthetic persona. Second, we find that group comparisons improve only with the respondent’s own record, and that even then the models’ group differences rest more on the demographic label in the prompt than on the individual respondent, which the usual check of reproducing known group differences does not reveal. Third, we find that the models make groups look more alike than they are, keeping differences in circumstances while flattening or reversing differences in attitudes, most sharply for religious affiliation and ethnic origin, both special categories of personal data. Fourth, we outline an audit that a survey organization can run on questions it has already fielded to decide which estimates may use the record. More broadly, our results argue for testing disclosure estimate by estimate wherever personal records are passed to a model that stands in for the people they describe. We release our code at \url{https://github.com/aspenywang/necessary-for-what}.
\end{abstract}
 

\renewcommand{\shortauthors}{Wang et al.}

\maketitle

\section{Introduction}
\label{sec:intro}

In 2026, Pew Research Center built ``digital twins'' of its survey panelists by giving a language model each panelist's demographic profile and earlier answers and asking it to complete new questionnaires as that person~\cite{chapekis2026}. Market-research firms and startups now sell synthetic panels built from respondents' data~\cite{toluna2025personas,qualtrics2026synthetic,simile2026}, and researchers use the design to fill in skipped questions and to anticipate answers to new ones~\cite{kim2024aiaugmented,park2024generative,toubia2025twin}. In each case, the demographic record that survey researchers collect to weight samples and compare groups~\cite{kish1965survey} is also placed in the prompt, on the premise that a richer profile makes a more faithful respondent~\cite{argyle2023outofone,santurkar2023whose,hwang2023aligning,batzner2025personae}. Comparing the twins with the panel by party required knowing each panelist’s party, but not necessarily telling the model that answered for them.

The distinction matters because the record describes a real person. Passing it to a model is a new use of personal data~\cite{andrus2022}, and ordinary attributes in combination can single someone out~\cite{sweeney2000simple,rocher2019estimating}. In our Pew sample, four demographic fields identify 0.1\% of respondents uniquely and the eight earlier answers in a prompt identify 18\%, but together they identify 90\%. Data minimization limits personal data to what is necessary for a purpose~\cite{gdpr,king2023privacybias,ganesh2025}, yet a synthetic survey serves several purposes at once: predicting how individuals answer, estimating how opinions are distributed, and comparing groups. A record that helps with one may add nothing to another. The question is not whether the record is necessary, but what it is necessary for.
\begin{figure}[t]
\centering
  \resizebox{0.75\textwidth}{!}{\begingroup%
\definecolor{figline}{HTML}{777777}%
\definecolor{figsep}{HTML}{D9D9D9}%
\definecolor{fixedblue}{HTML}{0072B2}%
\definecolor{variedorange}{HTML}{D55E00}%
\begin{tikzpicture}[
  x=1pt,y=1pt,inner sep=0pt,font=\fontsize{8}{9}\selectfont,
  line/.style={draw=figline,line width=.55pt},
  flow/.style={line,-{Stealth[length=3.2pt,width=2.8pt]}},
  fixedflow/.style={flow,draw=fixedblue},
  variedflow/.style={flow,draw=variedorange,dash pattern=on 2pt off 1.4pt},
  frame/.style={line,rounded corners=1.5pt},
  title/.style={anchor=base west,font=\fontsize{8.5}{10}\selectfont\bfseries},
  label/.style={anchor=base west,font=\fontsize{8}{9}\selectfont\bfseries},
  row/.style={anchor=base west},
  code/.style={anchor=base west,font=\ttfamily\fontsize{8}{9}\selectfont},
]

\draw[frame,fill=fixedblue!5] (0,0) rectangle (124,-110);
\draw[figsep,line width=.45pt] (0,-51)--(124,-51);
\node[title] at (0,5) {Respondent's record};
\node[label] at (6,-10) {Earlier answers ($H_i$)};
\node[code] at (6,-20) {Savings account? Yes};
\node[code] at (6,-29) {Government responsible for};
\node[code] at (6,-38) {healthcare? No};
\node[row] at (6,-47) {$\ldots$ 6 more question--answer pairs};
\node[label] at (6,-61) {Demographics ($D_i$)};
\node[code] at (6,-70) {Age: 30-49};
\node[code] at (6,-79) {Education: College graduate};
\node[code] at (6,-88) {Religion: Nothing in};
\node[code] at (6,-97) {particular};
\node[row] at (6,-106) {$\ldots$ 12 other fields};

\fill[fixedblue!5] (144,0) rectangle (266,-29);
\fill[variedorange!11] (144,-29) rectangle (266,-62);
\draw[frame] (144,0) rectangle (266,-134);
\draw[figsep,line width=.45pt] (144,-29)--(266,-29)
                                    (144,-62)--(266,-62);
\node[title] at (144,5) {Prompt};
\node[label] at (150,-10) {Earlier answers ($H_i$) --- fixed};
\node[row] at (150,-22) {Same eight question--answer pairs};
\node[label] at (150,-40) {Profile input ($Z_i$) --- varied};
\node[code] at (150,-50) {Own (full/reduced) /};
\node[code] at (150,-59) {Shuffled / External / None};
\node[label] at (150,-73) {Target question ($q$)};
\node[anchor=north west,align=left,
      font=\ttfamily\fontsize{8}{9}\selectfont] at (150,-77)
  {Thinking about the nation's\\
   economy, how would you rate\\
   economic conditions in this\\
   country today?};
\node[code] at (150,-120) {Excellent / Good /};
\node[code] at (150,-129) {Only fair / Poor};

\draw[frame] (272,-31) rectangle (292,-53);
\node at (282,-42) {LLM};
\draw[frame] (298,-25) rectangle (340,-59);
\node[align=center] at (319,-42)
  {Response\\[-1pt]probabilities\\[-1pt]$\hat p_{iq}$};

\draw[frame,draw=fixedblue,fill=fixedblue!5]
  (0,-114) rectangle (124,-138);
\node[align=center] at (62,-126)
  {Observed answer ($y_{iq}$)\\[1pt]Held out $\rightarrow$ evaluation only};

\fill[fixedblue!5] (346,-28) rectangle (430,-114);
\draw[frame] (346,0) rectangle (430,-114);
\draw[figsep,line width=.45pt] (346,-28)--(430,-28)
                                    (346,-55)--(430,-55)
                                    (346,-78)--(430,-78);
\node[title] at (346,5) {Evaluation};
\node[anchor=west] at (352,-14) {Individual predictions};
\node[anchor=west,align=left] at (352,-41.5)
  {Population/subgroup\\[-1pt]estimates};
\node[anchor=west] at (352,-66.5) {Group comparisons};
\node[label,text=fixedblue] at (352,-89) {Analysis groups};
\node[row,text=fixedblue] at (352,-99) {Age: 30--49 vs. 50--64};
\node[row,text=fixedblue] at (352,-109) {College / high school};

\draw[fixedflow] (124,-18)--(144,-18);

\draw[line,draw=fixedblue] (124,-76)--(134,-76);
\fill[fixedblue] (134,-76) circle (.9pt);
\draw[variedflow] (134,-76)--(134,-45)--(144,-45);
\node[fill=white,inner sep=.6pt,text=variedorange!80!black]
  at (134,-61) {Own};

\draw[flow] (266,-42)--(272,-42);
\draw[flow] (292,-42)--(298,-42);
\draw[flow] (340,-42)--(346,-42);

\draw[line,draw=fixedblue] (134,-76)--(134,-146);
\draw[line,draw=fixedblue] (62,-138)--(62,-146)--(134,-146);
\fill[fixedblue] (134,-146) circle (.9pt);
\draw[fixedflow] (134,-146)--(414,-146)--(414,-114);
\end{tikzpicture}%
\endgroup%
}
  \caption{Study design. For each respondent and target question, the model
  receives earlier answers ($H_i$), a target question ($q$), and a demographic
  input ($Z_i$) that varies across conditions. Own-profile conditions use the
  respondent's recorded demographics ($D_i$). Other conditions use a randomly chosen respondent's profile, an external profile, or none. The observed answer
  ($y_{iq}$) is withheld from the prompt and used for evaluation. In all main
  evaluations, group membership is defined by recorded $D_i$, regardless of
  what the model receives. Blue indicates fixed respondent information and orange indicates varied demographic input. Examples are illustrative.}
  \label{fig:study-design}
\end{figure}
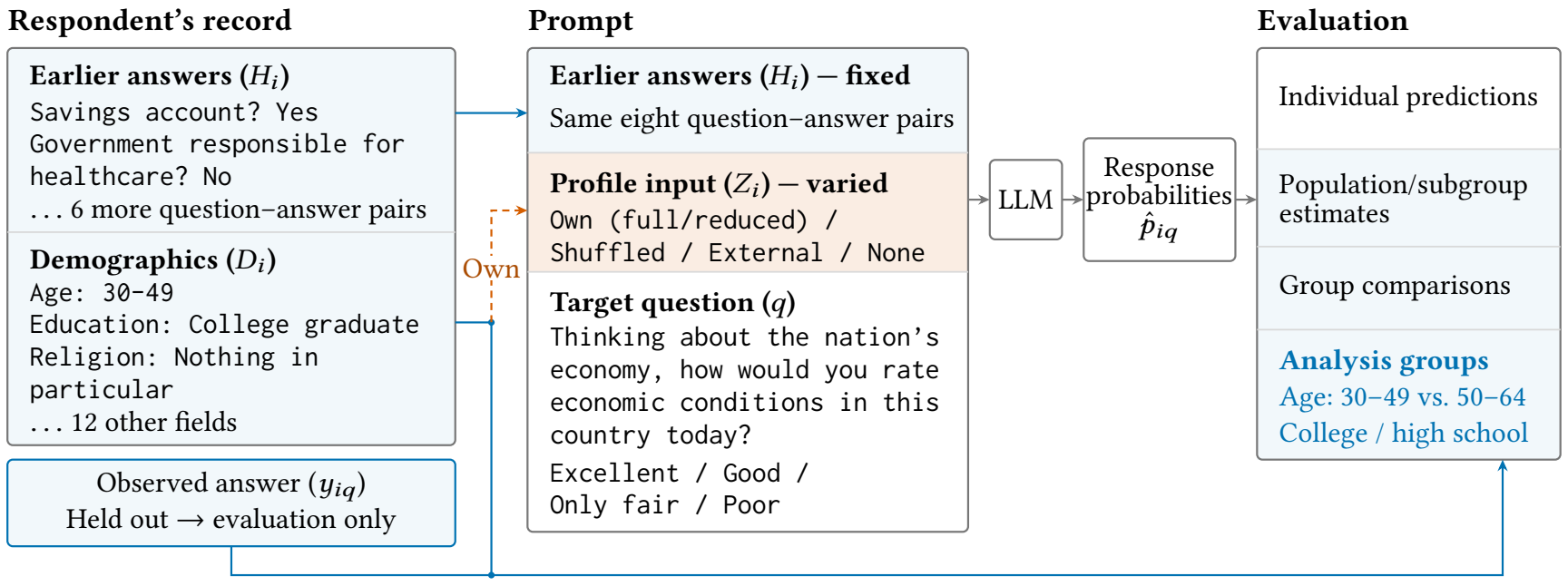

Evaluations of synthetic respondents have found that they can approximate aggregate opinion while misrepresenting differences between groups~\cite{bisbee2024synthetic,santurkar2023whose,wang2025misportray}, and that a respondent's earlier answers predict individual responses better than demographics do~\cite{park2024generative,pourtaheri2026behavioralhistory}. We ask how a respondent's record affects individual answers, population and subgroup estimates, and differences between groups (RQ1). Any gain raises two further questions. Does it require the respondent's own record, or would any plausible profile serve as well (RQ2)? And do improved estimates support sound conclusions about groups (RQ3)? The third matters because synthetic panels are often validated by reproducing known group differences~\cite{argyle2023outofone,chapekis2026}, and a model can pass this check by label-following, answering more from the demographic label in its prompt than from the individual respondent~\cite{cheng2023compost,peng2026}. We prompt models with each respondent's answers to other questions and add the respondent's own record, a randomly chosen respondent's record, a census record~\cite{census2019pums} or a synthetic persona~\cite{nvidia/Nemotron-Personas-USA}, with all of its fields or fewer (Figure~\ref{fig:study-design}). These variations mirror swapping, synthetic substitution and suppression in statistical disclosure control. We apply the design to the Pew American Trends Panel~\cite{pew2019atpw54,santurkar2023whose} and the International Social Survey Programme~\cite{issp2023}, with eleven open-weight models on both surveys and three hosted models on the Pew panel, and compare each with a statistical model fit to human responses with the same inputs.

We make four contributions. First, we show that whether the record is necessary depends on the estimate. It adds little to individual predictions, and for the open-weight models most of its gain in population and subgroup estimates survives replacing it with a census record or a synthetic persona. Group comparisons gain the most from it, and only when it belongs to the respondent. Second, we provide a regrouping test that groups answers produced from randomly assigned profiles by the profile in each prompt, and find that the models' group differences come mainly from label-following, a pattern a synthetic panel can show while still reproducing known group differences (Section~\ref{sec:linkage}). Third, we show whose views these errors misstate. Even with the respondent's own record, the models make groups look more alike than they are. They keep differences in circumstances such as savings and debts but flatten or reverse differences in attitudes toward government and inequality, and these errors persist in larger models. They are sharpest for religious affiliation and ethnic origin, two special categories of personal data, so for these respondents disclosure passes their most protected attributes and returns among the least faithful accounts of their views. Fourth, we outline an audit that a survey organization can run on questions it has already fielded to decide which estimates may use the record (Section~\ref{sec:disc-practice}).

Debates about demographic data in machine learning have centered on whether such data should be collected~\cite{andrus2021measure,bogen2020awareness,king2023privacybias}. In survey extension, the survey has already collected the records, so the question is which estimates justify passing them to a model, and whether conclusions drawn from those estimates hold. We answer the first question estimate by estimate and show that the answer does not settle the second.

\section{Background and Related Work}
\label{sec:background}
\label{sec:related}

\subsection{Synthetic Surveys and What They Are Evaluated On}
\label{sec:roles}

\paragraph{From simulated groups to extended panels.}
Language models have been prompted with demographic descriptions to simulate respondents of a given type, whose answers are compared with those of the corresponding group~\cite{argyle2023outofone,aher2023using,santurkar2023whose,li2024steerability,batzner2025personae,sun2024random}. A growing body of work instead extends panels of real respondents. A model receives what each panelist has already answered, often together with their demographic profile, and imputes skipped answers or anticipates answers to new questions~\cite{kim2024aiaugmented,park2024generative,hwang2023aligning,toubia2025twin,chapekis2026}. In survey extension, every synthetic respondent stands in for a real person whose record the researcher already holds, and demographics serve a further role. They define the groups for which results are reported, as when Pew evaluates its synthetic respondents by race, age and party~\cite{chapekis2026}.

\paragraph{Evaluating synthetic respondents.}
Evaluations have found that synthetic respondents can approximate aggregate opinion while misrepresenting how opinions vary across people and differ between groups~\cite{bisbee2024synthetic,qu2024performance,dominguezolmedo2024questioning,morocho2026reliability,santurkar2023whose,wang2025misportray,sen2025margins}. Models also lean on demographic identity more than real opinions warrant, producing caricatured or stereotyped portraits of groups~\cite{cheng2023compost,chen2026syntheticusers,chapekis2026}, and digital twins built from hundreds of a person's answers respond much like twins given demographics alone~\cite{peng2026}. A synthetic survey that gets a group comparison wrong misstates what a group of people think, which work on algorithmic harms treats as a representational harm~\cite{shelby2023harms,wang2025misportray}. We return to whose views are misstated in Section~\ref{sec:disc-groups}. A common validation criterion is whether a model reproduces known relationships between demographic attributes and responses~\cite{argyle2023outofone,bisbee2024synthetic,chapekis2026,lukauskas2026}, yet a model that answers from the group label alone can meet it. These studies measure how far synthetic groups depart from real ones. We test where synthetic group differences come from by regrouping answers produced from reassigned profiles by the profile shown (Section~\ref{sec:measures}).

\paragraph{Imputation and the relationships an analysis needs.}
Extending a survey with a model is a form of imputation, and missing-data research separates producing plausible values from supporting valid inference. Multiple imputation carries uncertainty about the missing values into the analysis~\cite{rubin1987multiple}, and an imputation model that omits or misrepresents relationships the analysis estimates can bias it even when the imputed values look plausible~\cite{meng1994multiple,bartlett2015accommodating}. Corrections that combine model predictions with a small set of human answers can restore valid population estimates~\cite{krsteski2026}. In fairness research, whether imputing a demographic attribute is legitimate has been separated from whether it yields valid group measurements~\cite{dong2026imputation}. We carry this kind of distinction to synthetic surveys. Whether a model receives the attribute that defines a group, and whether its responses preserve how that attribute relates to people's answers, are separate questions, and Section~\ref{sec:disc-profile} considers what this means for validating synthetic respondents.

\subsection{Demographic Information and Data Minimization}
\label{sec:design-space}

\paragraph{Measuring what demographics contribute.}
Demographic variables explain little of the variation in individual responses~\cite{hu2024quantifying}, and a respondent's earlier answers or interviews predict what they will say better than a demographic description does~\cite{park2024generative,zhang2025interview,pourtaheri2026behavioralhistory}. Adding demographics to earlier answers improves individual prediction only slightly~\cite{malone2026}. Evaluations usually estimate what demographics contribute by adding or removing them, or by varying which fields a profile contains~\cite{hwang2023aligning,bisbee2024synthetic,lukauskas2026,chapekis2026}. Replacement controls separate the information a profile carries from the presence of a profile. They replace survey-country labels~\cite{lyu2026}, match LLM-generated backstories to respondents at random~\cite{moon2024anthology} or swap single fields~\cite{lukauskas2026}, but none replaces a respondent’s full record. Profiles from public microdata or persona generators have been used to build synthetic populations~\cite{castricato2025persona,sun2024random}, but to our knowledge no study has compared such profiles with the records of the respondents a model stands in for. We reassign respondents' full recorded profiles while holding their own answers fixed, which lets us regroup the resulting answers by the profile in the prompt, and we compare respondents' records with census and persona profiles. Together, these separate the effect of supplying demographic information from the effect of linking it to the respondent, for each estimate a survey reports.

\paragraph{Demographic data and data minimization.}
Work on fairness in machine learning has examined the risks of collecting the demographic data needed to measure disparities~\cite{bogen2020awareness,andrus2021measure,andrus2022}, and minimization can pull against that measurement~\cite{king2023privacybias}. Data minimization has been operationalized for personalization~\cite{biega2020minimization}, as a constraint on the features a model may use~\cite{ganesh2025} and on the inputs it receives at inference time~\cite{tran2023data}. In synthetic survey research, privacy has been treated mainly as a matter of data handling, through consent, de-identification, data-use agreements, local models and the omission of direct identifiers~\cite{lukauskas2026,garzon2026surveyanchors,rupprecht2026,lyu2026,frohling2026personaattributes}, and some authors note the risk of disclosing more sensitive information as personas grow richer~\cite{frohling2026personaattributes} or of inferring people's beliefs~\cite{malone2026}. We apply minimization to the prompt itself. In survey extension the researcher holds each record whatever the model receives, so the question is whether passing the record to the model is necessary for the estimate the response serves, which Section~\ref{sec:disc-necessity} answers estimate by estimate.

\section{Experimental Design}
\label{sec:design}

\subsection{Study Setting}
\label{sec:setting}

We evaluate synthetic respondents on completed surveys, where a respondent's answers to some questions are used to estimate their answers to others (Section~\ref{sec:roles}). For respondent $i$ and target question $q$, the model receives a history of the respondent's answers to other questions $H_{iq}$ and a demographic input $Z_i$, and returns a distribution $\hat p_{iq}$ over the response options $k$. Because every target was answered, with recorded answer $y_{iq}$, each prediction, and each survey estimate built from predictions, can be compared with its human counterpart (Figure~\ref{fig:study-design}). Across conditions, we vary $Z_i$ while holding the history, the target question and the model fixed. The researcher always retains each respondent's recorded profile $D_i$ and uses it to define the groups in every subgroup and group-comparison estimate, whatever profile the model received.

We use three terms throughout. A respondent's \emph{record} is the demographic profile $D_i$ the survey recorded. A \emph{profile} is any demographic input $Z_i$ the model receives, whether the respondent's record, a randomly chosen respondent's, a census record or a synthetic persona, and prior work calls giving a model such an input sociodemographic or persona prompting~\cite{beck2024sociodemographic,hu2024quantifying,lutz2025prompt}. Following work on data minimization at inference time~\cite{tran2023data}, \emph{disclosure} is passing the record, or some of its fields, to the model. This holds whether the model runs on the researcher's own machines, as our open-weight models do, or is operated by a third party, which also discloses the record to that party~\cite{gdpr}. Survey records are collected to weight samples and define groups, and disclosure puts them to a further use, shaping the answers a model gives in the respondent's place~\cite{andrus2022}.

\subsection{Data and Models}

\paragraph{Surveys.} We use two surveys of U.S. respondents on contested public issues. Wave 54 of Pew Research Center's American Trends Panel, distributed with OpinionQA~\cite{pew2019atpw54,santurkar2023whose} (the Pew panel), covers economic inequality, and we use 1,000 respondents and 63 target questions. The 2023 International Social Survey Programme module on National Identity and Citizenship~\cite{issp2023} covers national pride, immigration, citizenship and trust in institutions, and we use 1,513 respondents and 27 target questions. In each survey, the record has 15 demographic fields (Appendix Table~\ref{tab:fields}). For each target, the history consists of eight of the respondent's answers to other questions, chosen by a fixed rule that does not depend on answer values (Appendix~\ref{app:history-rule}). For seven models we also vary the length of the history, down to no history at all (Appendix~\ref{app:history}).

\paragraph{External profiles.} Alternative profiles come from the 2019 American Community Survey Public Use Microdata Sample~\cite{census2019pums}, matching the year of the Pew wave, and from Nemotron-Personas-USA, a set of synthetic personas generated to match U.S. census distributions~\cite{nvidia/Nemotron-Personas-USA}. Ten Pew fields have ACS counterparts, four of which also have Nemotron counterparts. External values are mapped to the survey's categories (Appendix~\ref{app:external}).

\paragraph{Models.} We evaluate eleven open-weight, instruction-tuned models from six families, with 7B to 72B parameters: Gemma~4~\cite{gemmateam2026gemma4}, OLMo~3~\cite{olmo2025olmo3}, Qwen2.5~\cite{qwen2024qwen25}, Qwen3.5~\cite{qwen2026qwen35}, Llama~3.1~\cite{grattafiori2024llama3} and Mistral~7B~\cite{jiang2023mistral} (Appendix Table~\ref{tab:models}). Open weights let us score every response option and keep the restricted ISSP records on local machines. Each model's response distribution normalizes the likelihoods of the complete answer strings for the valid options. On the Pew panel, which is public, we also evaluate three hosted models, GPT-4.1~\cite{openai2025gpt41}, GPT-6.1 Terra~\cite{openai2026gpt61terra} and Claude Haiku 5.5~\cite{anthropic2026haiku55}, and describe their scoring in Appendix~\ref{app:models} and report them in Appendix~\ref{app:hosted}.

\subsection{Comparisons}
\label{sec:strategies}

The conditions vary the profile along three dimensions (Table~\ref{tab:comparisons}). \emph{Linkage} is whether the profile describes the respondent, \emph{provenance} is whether it comes from the respondent's record, public microdata or a synthetic generator, and \emph{granularity} is how many attributes it contains. Each dimension corresponds to a tool of statistical disclosure control~\cite{hundepool2012sdc}, namely record swapping~\cite{dalenius1982swapping}, released or synthetic microdata~\cite{rubin1993discussion} and field suppression~\cite{sweeney2002kanon}. \textsc{Shuffled} reassigns the recorded profiles among respondents with one fixed random permutation and serves as a control for linkage. Census and persona profiles are assigned independently of the respondent. Reduced profiles drop fields along one fixed suppression path~\cite{sweeney2002achieving}, so they measure the cost of cumulative reduction. These comparisons make necessity testable. We call the record \textit{necessary} for an estimate when \textsc{Own} improves it over \textsc{History} and over every profile that does not describe the respondent, and a field \textit{necessary} when removing it worsens the estimate.

\begin{table}[th]
\caption{Conditions. Every condition includes the respondent's eight-answer history and uses the same prompt, so only the profile block differs (Appendix~\ref{app:prompts}). Substitution and minimization conditions use the Pew panel, and all conditions use the eleven main models.}
\label{tab:comparisons}
\small
\begin{tabular}{@{}l p{0.34\linewidth} l p{0.30\linewidth}@{}}
\toprule
Condition & Profile the model receives & Compared with & Question \\
\midrule
\multicolumn{4}{@{}l}{\textit{Value and linkage}}\\
\textsc{History} & None (country only) & -- & Baseline \\
\textsc{Own} & The respondent's 15 recorded fields & \textsc{History} & Does the respondent's record help? \\
\textsc{Shuffled} & A random respondent's 15 fields & \textsc{Own} & Does correct linkage matter? \\
\addlinespace
\multicolumn{4}{@{}l}{\textit{Substitution}}\\
\textsc{Shuffled-10} & A random respondent's same 10 fields & \textsc{Own-10} & Does linkage matter on these fields? \\
\textsc{PUMS-10} & A census person record with the same 10 fields & \textsc{Own-10} & Can public microdata substitute? \\
\textsc{Nemotron-4} & A synthetic persona with 4 fields & \textsc{Own-4} & Can synthetic personas substitute? \\
\textsc{Nemotron-All} & The same persona with 9 structured fields & \textsc{Nemotron-4} & What does a richer persona add? \\
\addlinespace
\multicolumn{4}{@{}l}{\textit{Minimization}}\\
\textsc{Own-10/7/4} & Subsets of the respondent's fields (Appendix Table~\ref{tab:fields}) & \textsc{Own} & What do fewer fields cost? \\
\bottomrule
\end{tabular}
\end{table}

\subsection{Measures}
\label{sec:measures}

All conditions are scored on the same respondent--question cells with respondents weighted equally, and groups are defined by respondents' recorded attributes, except in the regrouping test below. 

\textit{Fidelity.}
We call the agreement between synthetic and human responses \emph{fidelity} and measure it at three levels: individual answers, the whole sample and each subgroup. For individual answers we use the multiclass Brier score $\sum_k (\hat p_{iqk} - \mathbb{1}[y_{iq}=k])^2$, a strictly proper scoring rule~\cite{gneiting2007strictly}. For a group $g$, which is the whole sample for population estimates, let $\hat p_{gqk}$ be the mean of $\hat p_{iqk}$ over its members and $p_{gqk}$ the share of its members who chose $k$. We compare the two distributions with total variation distance, $100 \cdot \tfrac{1}{2}\sum_k |\hat p_{gqk} - p_{gqk}|$ in percentage points~\cite{meister2025benchmarking}, averaged over target questions and, for subgroups, over groups.

\textit{Group differences.}
For each target question we fix a side $S_q$ of the response scale, and define the contrast between groups $g$ and $h$ of the same attribute as $c(q, g, h) = 100 \sum_{k \in S_q} (\hat p_{gqk} - \hat p_{hqk})$, the difference in percentage points between the shares of the two groups on that side, with $c^{\mathrm{hum}}$ defined in the same way from human shares. A pair is \emph{human-supported} when $|c^{\mathrm{hum}}| \geq 5$ percentage points (pp) and its 95\% bootstrap interval (Section~\ref{sec:inference}) excludes zero (1,281 pairs in the Pew panel and 1,268 in the ISSP, all on attributes shown in the profile, Appendix~\ref{app:measures}). Applying the same rule to the synthetic contrast, we call a human-supported difference \emph{reproduced} when the synthetic contrast meets it with the human sign, \emph{reversed} when it meets it with the opposite sign, \emph{missed} otherwise, and \emph{resolved} when it is reproduced or reversed. A missed difference is synthetic-null when its 90\% interval lies within $\pm 5$~pp. We also report the contrast error $|c - c^{\mathrm{hum}}|$, averaged over human-supported pairs, and the ratio $c / c^{\mathrm{hum}}$, summarized by its median over pairs.

Group-level measures are reported for each attribute separately, because attributes contribute very different numbers of comparisons and their groups mean different things~\cite{wang2025identities}. To see where group differences go when profiles are reassigned, we also group \textsc{Shuffled} predictions by the profile shown in the prompt and apply the same grouping to \textsc{History} predictions, which saw no profile, as a baseline. We call group differences label-following when more of them appear with this grouping than with groups defined by the respondents’ own records, that is, when they rest more on the demographic label in the prompt than on the individual respondent.

\textit{Identifiability.}
Combinations of ordinary attributes can narrow the set of people a profile could describe~\cite{sweeney2000simple,rocher2019estimating,shlomo2022measuring}. For 4, 7 and 10 fields, we estimate how many U.S.\ adults share each respondent's values, from person weights of exact matches in the 2019 ACS PUMS. We also count how many respondents are unique in the public Wave 54 file on their profile fields, alone and with the answers in their prompt (Appendix~\ref{app:identifiability}).

\subsection{Reference Model and Inference}
\label{sec:inference}

To put the models' gains in context, we compare each with a statistical reference that receives the same history and profile fields. The reference is an L2-regularized multinomial logistic regression for each target question and condition, trained with five-fold cross-fitting~\cite{chernozhukov2018double} on other panel respondents' answers to that question (Appendix~\ref{app:inference}). It shows how much the same information improves predictions when a model is fit to human data, and because it needs answers to the target question, it is available only for questions already fielded. For a measure with error $L$, the gain of condition $Z$ over condition $C$ is the paired reduction in error, $L(C) - L(Z)$, so a positive gain means lower error under $Z$. We estimate uncertainty with 10,000 paired bootstrap resamples of respondents, shared across all models and conditions, and report 95\% confidence intervals. A comparison improves or worsens fidelity for a model when its interval excludes zero, and is unresolved otherwise. Across the eleven main models we report the median effect, its range and the number of models in each category, alongside the reference.

\section{What the Respondent's Record Adds}
\label{sec:results}

We compare conditions by how far their estimates fall from the human ones, for individual predictions, population and subgroup distributions, and group comparisons. Section~\ref{sec:value} measures what the respondent's own profile adds (RQ1), and Sections~\ref{sec:linkage} and~\ref{sec:alternatives} ask whether that gain requires the respondent's record (RQ2). Aggregate estimates gain little, and much of that gain does not depend on whose profile the model receives. Group comparisons gain more, but only from the respondent's own profile. Appendix~\ref{app:hosted} reports the three hosted models.

\subsection{The Value of Disclosure Depends on the Estimate}
\label{sec:value}

\begin{figure*}[t]
\centering
\includegraphics[width=0.7\textwidth]{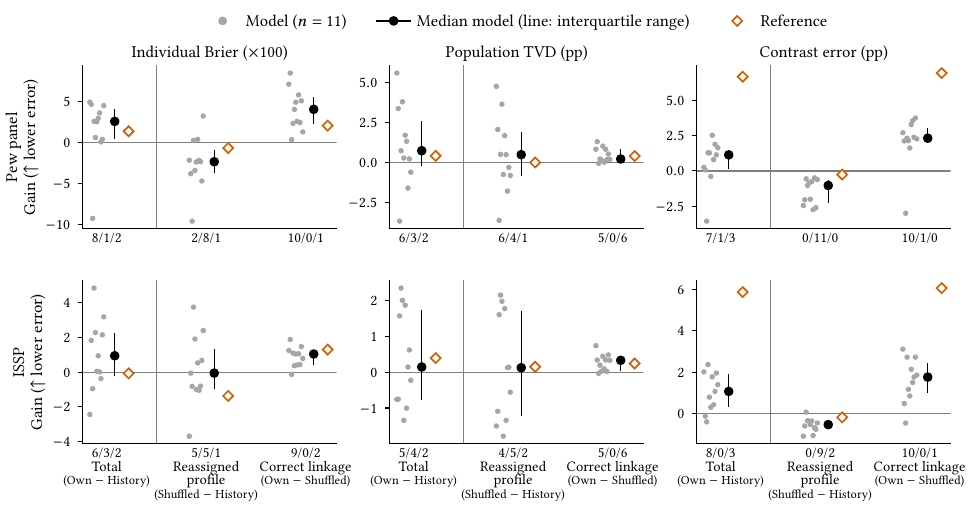}
\caption{For the median model, individual predictions and group comparisons improve with the respondent's own profile but not with a reassigned one, while a reassigned profile moves population estimates about as much as the respondent's own. The columns compare \textsc{Own} with \textsc{History}, \textsc{Shuffled} with \textsc{History}, and \textsc{Own} with \textsc{Shuffled}. Positive values indicate lower error under the first condition. Grey dots are the eleven main models, the black dot their median with its interquartile range, and the orange diamond the statistical reference. Counts below each category give the number of models that improve, worsen or are unresolved. Brier gains are $\times 100$, other gains in pp.}
\label{fig:gains}
\end{figure*}

Individual predictions improve only slightly. As Figure~\ref{fig:gains} shows, \textsc{Own} lowers the median model's Brier score by 3\% in the Pew panel and 1\% in the ISSP, a little more than it lowers the reference's. Once a model has seen eight of a respondent's answers, their demographics add little, and they add less the more answers the model sees (Appendix~\ref{app:history}).

Population estimates barely move. With answers alone, the models' population estimates lie 22 to 51~pp (TVD) from the surveyed sample, against about 2~pp for the reference. Adding the profile narrows this distance by less than 1~pp for the median model.

Group comparisons are where the profile helps most, and where the models fall furthest behind the reference. Contrast error falls for 7 of 11 models in the Pew panel and 8 of 11 in the ISSP, and in the Pew panel the median share of human-supported differences the models reproduce rises from 22.6\% to 41.8\%. The models start almost as close to the human contrasts as the reference does, yet the reference gains five to six times as much from the same profile.

\subsection{Linkage Matters Differently Across Estimates}
\label{sec:linkage}

\begin{figure*}[th]
\centering
\includegraphics[width=0.7\textwidth]{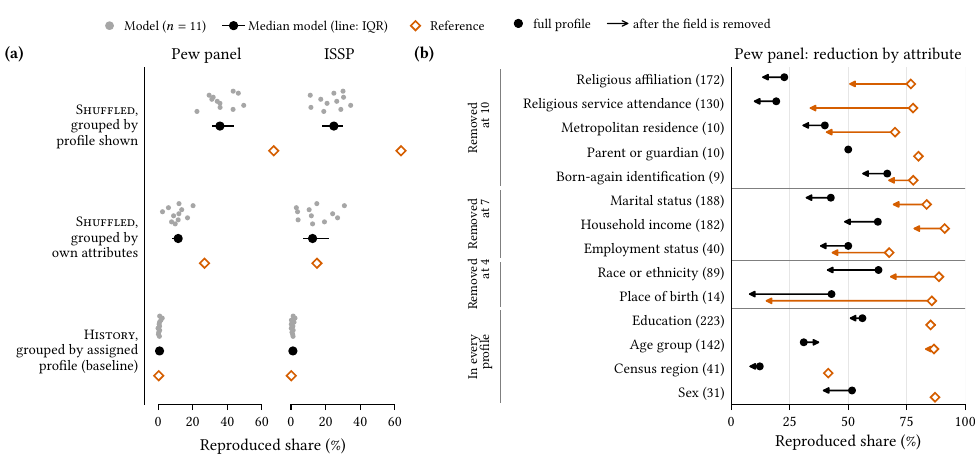}
\caption{The models' group differences follow the label in the prompt, whether it is reassigned (a) or removed (b). (a) Each respondent's own answers are paired with another respondent's profile (\textsc{Shuffled}), and the predictions are grouped either by the profile shown in the prompt or by the respondents' own attributes. Each row gives the share of human-supported differences that grouping reproduces. \textsc{History} predictions, which saw no profile, grouped by the assigned profiles serve as a baseline. Grey dots are the eleven main models, black is their median with its interquartile range, and orange is the statistical reference fit to respondents' own records and given the same reassigned profiles. (b) Share of each attribute's human-supported differences reproduced with the full profile (dot) and after the attribute is removed along the reduction path (arrowhead), for the median model (black) and the reference (orange), Pew panel, with pair counts in parentheses. Attributes kept in every profile are shown at four fields.}
\label{fig:assigned}
\end{figure*}

Does the gain require the respondent's own profile? For population estimates, mostly not. As Figure~\ref{fig:gains} shows, \textsc{Shuffled} moves the models' population estimates about as much as \textsc{Own}, which adds a further gain for only five models in each survey. Much of what a profile does to population estimates comes from receiving one at all, with GPT-6.1 Terra as an exception (\textsc{Shuffled} recovers only a third of its gain from \textsc{Own}, Appendix~\ref{app:hosted}). The reference, in contrast, gains almost nothing from \textsc{Shuffled}. For group comparisons, another respondent's profile is worse than no profile at all. \textsc{Shuffled} makes contrast error worse for all 11 models in the Pew panel and 9 of 11 in the ISSP, and in the Pew panel it cuts the median share of reproduced differences from 22.6\% to 11.6\%. The same split holds with no earlier answers, where the profile is the model's only information about the respondent. Here (Appendix~\ref{app:history}), \textsc{Own} and \textsc{Shuffled} each improve population estimates over \textsc{History} by 8 to 10~pp. For group comparisons, \textsc{Own} reproduces 40\% of human-supported differences in the Pew panel and 34\% in the ISSP, about as many as with eight answers, while \textsc{Shuffled} reproduces about 1\%.

The models’ group differences follow the profile in the prompt. When we group the \textsc{Shuffled} predictions by the profile shown in each prompt (Figure~\ref{fig:assigned}a), they reproduce 36\% of human-supported differences in the Pew panel and 25\% in the ISSP, against 12\% in both surveys when grouped by respondents’ own attributes. Most are the differences \textsc{Own} reproduces. Of those, 80\% in the Pew panel and 57\% in the ISSP are also reproduced from another respondent’s profile when grouped by that profile. The same grouping applied to \textsc{History} predictions, which never saw a profile, reproduces about 1\%. Household income shows the pattern clearly. Respondents with incomes of \$100,000 or more are 45~pp more likely than those earning \$30,000--50,000 to hold stocks, bonds or mutual funds outside retirement accounts. In the median model, respondents shown these two income profiles differ by 51~pp, while the same predictions grouped by respondents’ actual incomes differ by 2~pp. The models’ group differences are therefore largely label-following, more so in the Pew panel than in the ISSP. A reference fit to respondents’ own records also follows the label when given the reassigned profiles (Figure~\ref{fig:assigned}a). It draws more on the respondents, however, reproducing 27\% and 15\% when grouped by their own attributes, against 12\% for the median model in both surveys.

\subsection{Disclosing Less Costs Group Comparisons}
\label{sec:alternatives}

A researcher could disclose less by changing a profile's provenance, replacing the respondent's record with a census or persona profile, or by reducing its granularity, giving the model fewer of the respondent's attributes. Replacement benefits aggregates and hurts group comparisons. For the median model, census and persona profiles with the same fields improve population and subgroup estimates by about as much as the respondent's own record (Appendix~\ref{app:alternatives}). With them, however, most models predict individual answers less accurately than with answers alone, no model improves contrast error, and the models reproduce fewer human-supported differences than with no profile (13\% and 18\%, against 23\%). A richer persona does not help. \textsc{Nemotron-All} gives the model the same persona as \textsc{Nemotron-4} with nine structured fields instead of four, adding details such as exact age, occupation and city, yet population and subgroup errors rise above those with no profile, in line with evidence that richer generated personas can move simulated opinions away from human ones~\cite{ahn2026,li2025persona}.

Reduction costs less, but the cost again falls on group comparisons. Along the reduction path, the models reproduce 41.8\% of human-supported differences with the full profile and 40.0\% with ten fields, while aggregate gains show no consistent trend. As Figure~\ref{fig:assigned}b shows, the loss falls on the comparisons defined by the removed fields. Attributes still in the profile lose at most 2.5~pp of reproduced share, against 7.6 to 14.0~pp for those removed. The ten-field profile, for example, drops religion, and the median model's reproduction of differences between religious groups falls from 23\% to 13\%. The remaining fields and answers still carry related information, since the reference reproduces half of these differences at ten fields. The cost of a smaller profile thus depends on which comparisons a study will report. Removing fields also does less for privacy than it seems. Reducing the profile from 15 fields to four lowers the share of respondents unique in the public Wave 54 file from 95\% to 0.1\%, but with the eight answers in a prompt, four fields still make 90\% unique (Table~\ref{tab:identifiability}).

\section{What Disclosure Does to Group Conclusions}
\label{sec:groups}

Survey reports state many findings as differences between groups~\cite{kish1965survey}, as Pew's evaluation of its synthetic respondents does~\cite{chapekis2026}, and a lower contrast error does not tell a researcher which of those differences a synthetic survey would get right (RQ3). We therefore follow each human-supported difference from \textsc{History} to \textsc{Own} and classify it as reproduced, reversed or missed, and call it resolved when it is reproduced or reversed (Figure~\ref{fig:transitions}). With the profile, the models resolve nearly twice as many differences, yet most remain missed and about one in nine of those they resolve still points the wrong way. These errors are largest for religion in the Pew panel and ethnic origin in the ISSP, larger models reduce them without closing the gap, and average fidelity does not reveal them.

\begin{figure*}[t]
\centering
\includegraphics[width=0.75\textwidth]{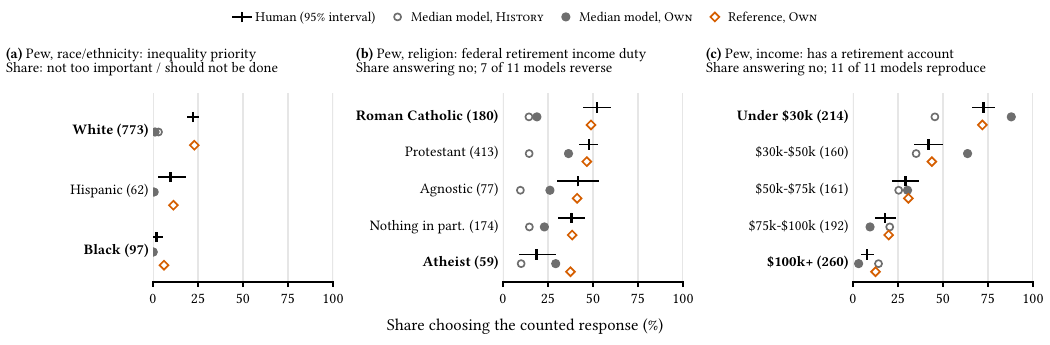}
\caption{With the respondent's profile, the models miss a difference between racial groups (a), reverse a difference between religious groups (b), and reproduce, even overstate, a difference by income (c). For every group of one attribute, each panel shows the share of human respondents choosing the counted response (with 95\% interval), the median model's share without the profile (\textsc{History}, hollow) and with it (\textsc{Own}, filled), and the reference with the profile (orange). The two groups compared in the text are in bold.}
\label{fig:examples}
\end{figure*}

\textbf{Models make groups look more alike than they are.}
Disclosure nearly doubles the share of human-supported differences the models reproduce, but 55\% remain missed. The differences the models do produce are mostly too small, which we call flattening. \citet{wang2025misportray} use the term for portraying a group as more uniform than its members, and we use it for differences between groups that shrink toward zero. The median model's differences are about a fifth as large as those respondents reported (median ratio 0.21 in both surveys), and for three in ten differences the models' responses would lead a researcher to conclude that the two groups differ by less than 5~pp. The reference, by comparison, misses 19\% of the differences, recovers more than half of their size (median ratio 0.74 in the Pew panel and 0.55 in the ISSP) and would lead to that conclusion for 8\% of them. Figure~\ref{fig:examples}a shows one case. Black respondents in the Pew panel are 20~pp less likely than White respondents to say that reducing economic inequality should not be a priority for the federal government. With race in the prompt, seven of the eleven models place the two groups within 5~pp of each other (the reference gives 17~pp). A study built on these responses would report little difference between Black and White respondents on this question.

\begin{figure*}[t]
\centering
\includegraphics[width=0.5\textwidth]{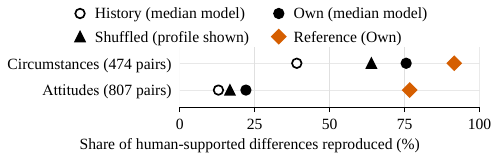}
\caption{Models reproduce more differences in circumstances than in attitudes. The Pew panel is split by question wording in an exploratory analysis. Black markers show median reproduced shares across the eleven open-weight models under \textsc{History} (hollow circles), \textsc{Own} (filled circles), and \textsc{Shuffled} grouped by the profile shown (triangles). Orange diamonds show the statistical reference under \textsc{Own}. Counts are human-supported group comparisons.}
\label{fig:circumstances}
\end{figure*}

\textbf{Differences in attitudes are flattened and reversed more than differences in circumstances.}
In an exploratory split of the Pew questions by wording into respondents' own circumstances, such as debts and savings,    and their attitudes, such as government responsibilities and the causes of inequality, a distinction long drawn in survey methodology~\cite{schaeffer2003science} (Appendix~\ref{app:circumstances}), the median model reproduces 76\% of the 474 differences in circumstances and 22\% of the 807 differences in attitudes. The reference reproduces 92\% and 77\% (Figure~\ref{fig:circumstances}). The gap remains among large differences. Of the differences of 20~pp or more, the median model recovers 92\% of those in circumstances and 34\% of those in attitudes, against 99\% and 94\% for the reference. Differences in attitudes are also reversed more often, for six of the seven attributes with enough differences of each type (Appendix~\ref{app:circumstances}). The profile alone produces much of what the models keep. Grouped by the profile shown, \textsc{Shuffled} predictions reproduce 64\% of the differences in circumstances and 17\% of those in attitudes.

\begin{figure*}[t]
\centering
\includegraphics[width=0.7\textwidth]{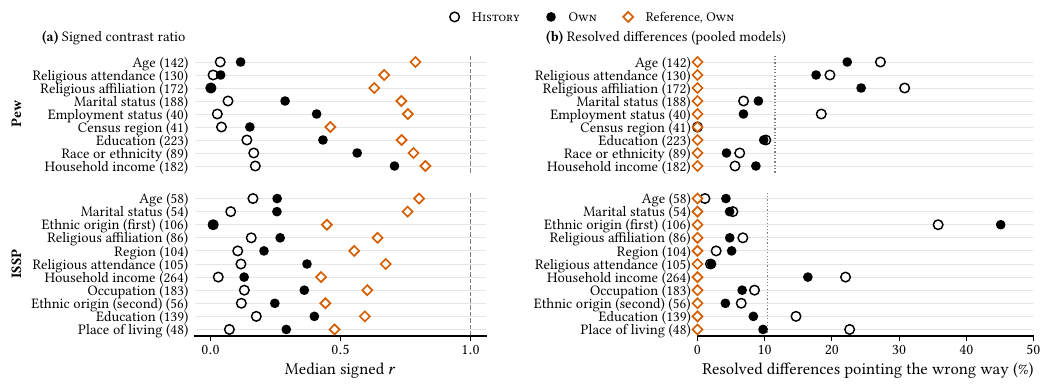}
\caption{Differences by religion and ethnic origin are flattened and reversed the most. (a) Median ratio of synthetic to human difference
over each attribute’s human-supported differences, for the median model without (hollow) and with (filled) the respondent’s profile
and for the reference (orange). The dashed line marks the human size. (b) Share of each attribute’s resolved differences that point the
wrong way, pooled over the eleven models, without (hollow) and with (filled) the respondent’s profile, and for the reference with
the profile (orange). The dotted line marks the share over all attributes in that survey with the profile. Attributes with at least 40
human-supported differences, with counts in parentheses.}
\label{fig:pew-question-types}
\end{figure*}

Religion and ethnic origin show the loss most sharply. For religious affiliation in the Pew panel, the median model keeps 0.63 of the size of differences in circumstances and none of the size of differences in attitudes (median ratios 0.63 and 0.00). In the ISSP, whose questions concern views on national identity, citizenship and trust, differences by first-mentioned ethnic origin stay near zero (0.01, Figure~\ref{fig:pew-question-types}a). The reference recovers 0.45 to 0.67 of these religious and ethnic differences. Over all attributes, about one in nine resolved differences point the wrong way, but about a quarter of those by religious affiliation in the Pew panel do, as do a fifth of those by age in the Pew panel and almost half of those by first-mentioned ethnic origin in the ISSP (pooled over models, Figure~\ref{fig:pew-question-types}b). In the Pew panel, Roman Catholic respondents are 34~pp more likely than atheist respondents to say that the federal government is not responsible for providing adequate income in retirement (Figure~\ref{fig:examples}b). With religion in the prompt, seven models place atheists above Catholics, the opposite of what the survey found. Differences by household income fare far better. They reach 0.71 of their human size, fewer than one in ten resolved income differences point the wrong way, and every model reproduces the difference in retirement accounts in Figure~\ref{fig:examples}c.

\textbf{These errors persist in larger models and in models with better average fidelity.}
In each of the five families we run at two sizes, the larger model reproduces 6 to 18~pp more of the Pew differences and, pooled over both surveys, newly reverses fewer of them (Appendix~\ref{app:size}). The best models, open-weight or hosted, still reproduce about 60\% of the Pew differences, against 82\% for the reference, and four of the five larger open-weight models reverse more differences with the profile than without it. Average fidelity does not single out the models that make these errors. Of the seven models whose contrast error improves with the profile in the Pew panel, five also reverse more differences with it, as do seven of eight in the ISSP. Llama~3.1~8B, for example, lowers its contrast error by 1.9~pp in the Pew panel while reversing 104 more differences.

\section{Discussion}
\label{sec:discussion}

Measuring disparities usually requires collecting demographic data~\cite{andrus2021measure,bogen2020awareness,king2023privacybias}. However, in survey extension, every group is defined by records the survey already collected, so the open question is which estimates should pass the record to the model. Our results answer this estimate by estimate. Population and subgroup estimates gain little from the respondent's record, and group comparisons need it (Section~\ref{sec:linkage}). Where the record is needed, the models use it mainly as a label for the group and take little from the person, and the group differences they produce keep what the label predicts and lose much of what people in each group think (Section~\ref{sec:groups}). We take up what this means for data minimization, for how synthetic respondents are validated, for the groups a synthetic survey describes and for how disclosure can be audited.

\subsection{Necessity Depends on the Estimate}
\label{sec:disc-necessity}

Data minimization judges necessity against a purpose~\cite{king2023privacybias,ganesh2025}, and in a synthetic survey the purpose has to be specified at the level of the estimate, because the answer differs across the estimates a single synthetic survey produces. For population and subgroup estimates, the respondent's record adds little, and for the open-weight models census and persona profiles serve about as well (Sections~\ref{sec:value} and~\ref{sec:alternatives}). These estimates can be produced without passing the record to the model. An improvement still has to be large enough for the study's purpose. Even with the record, most group differences are missed or flattened (Section~\ref{sec:groups}), so a case for disclosure also has to state how accurate a study needs its comparisons to be. The hosted models also show the trade-off this creates. GPT-4.1 and GPT-6.1 Terra gain about twice as much from the record in individual predictions as the median open-weight model, and a census profile does not substitute for it in their individual predictions or group comparisons (Appendix~\ref{app:hosted}), yet a hosted model receives the record as a third party. Group comparisons gain the most from the record, and only when the profile describes the respondent (Section~\ref{sec:linkage}).

For group comparisons, necessity also applies field by field. Every field passed to the model needs a reason, and our results tie that reason to the analysis plan, since removing a field costs mainly the comparisons it defines (Section~\ref{sec:alternatives}). A study can therefore limit the profile to the fields that define the groups it will compare. This matters most for special categories of personal data, such as religion and ethnic origin~\cite{gdpr}, which a study then passes only when it compares those groups. Fewer fields do much less to keep a respondent from being singled out. The answers in a prompt already make about one in five Pew respondents unique in the released panel~\cite{narayanan2008robust}, and four demographic fields beside them make nine in ten unique (Section~\ref{sec:alternatives}). For estimates that do not need the record, leaving it with the analyst is both the simplest minimization and the one that most limits who can be singled out.

This setting changes the privacy-bias tradeoff that has shaped debates about demographic data~\cite{king2023privacybias,andrus2021measure, bogen2020awareness}. That tradeoff arises because measuring a disparity requires knowing who belongs to which group. In survey extension the analyst knows this from the record already held, whatever the model receives, so withholding the record from the prompt costs little for population estimates and leaves every group comparison possible. The tradeoff returns only for group comparisons, and what the record adds there depends on how the models use it. 

\subsection{Validating Synthetic Respondents Requires Separating the Label from the Person}
\label{sec:disc-profile}

A synthetic respondent built from a real person's record and answers still responds largely as its demographic label suggests, much as invented personas do~\cite{cheng2023compost,wang2025misportray}. The models produce largely the same group differences from another respondent's profile, grouped by the profile shown, as from the respondent's own. Their group differences are label-following. A statistical model fit to the same data shows the same pattern, but it also draws more on what the respondents actually answered (Section~\ref{sec:linkage}).

Such distinctions matter for how synthetic surveys are validated. A common check that synthetic respondents represent people is whether they reproduce known relationships between demographics and responses \cite{argyle2023outofone,bisbee2024synthetic,chapekis2026,lukauskas2026}, and a model can pass much of this check from labels alone. Detecting this requires a control that separates the label from the person, such as reassigning profiles and grouping answers by respondents' own attributes. In missing-data terms, the models are uncongenial to the analysis~\cite{meng1994multiple}, filling in plausible answers without preserving the relationship between group membership and answers that a group comparison estimates~\cite{bartlett2015accommodating}. The flattened and reversed differences in Section 5 show this.

\subsection{Whose Views a Synthetic Survey Misstates}
\label{sec:disc-groups}

Differences in circumstances may survive label-following because they follow closely from income, age and education, while attitudes vary widely within every group. The models keep the first, largely from the label alone, and flatten and reverse the second (Section~\ref{sec:groups}). This fits evidence that digital twins given hundreds of a person's answers still respond much like twins given demographics alone, answering as a typical member of their group~\cite{peng2026}. The reference, which learns from respondents' answers, recovers most differences of both kinds, though fewer in attitudes.

These errors misrepresent groups. A synthetic survey that flattens differences in attitudes portrays groups as holding the same views when they differ, and one that reverses them changes which group a study portrays as more likely to endorse a position~\cite{shelby2023harms,wang2025misportray}. With six of the eleven models, for example, a study of national identity would find respondents of Mexican origin less likely than those of English origin to see American ancestry as important for being truly American, the reverse of what they reported. The reference reverses none of these differences, so such associations are not supported by the respondents' answers and are consistent with associations learned from text~\cite{cheng2023compost,wang2025misportray}.

The two groups with the largest errors are those that data protection treats with the most care. Religious beliefs and racial or ethnic origin are special categories of personal data, whose processing needs the strongest justification~\cite{gdpr}. Religion in the Pew panel and ethnic origin in the ISSP are where the models lose and reverse the most (Section~\ref{sec:groups}). For these respondents, passing the record discloses their most sensitive attributes and returns among the least faithful accounts of their views. The privacy cost and the representational error of disclosure fall on the same people.

\subsection{Auditing Disclosure by Analytic Purpose}
\label{sec:disc-practice}

A researcher extending a survey to new questions cannot fit a reference to answers that do not yet exist, so the gap we measure would go unseen unless disclosure is first tested on questions people have already answered~\cite{krsteski2026}. Such a test can be organized around the estimates a synthetic survey will report and how accurate each must be. Comparing answers alone with own and reassigned profiles shows what disclosure and correct linkage add. Regrouping reassigned predictions against a reference fit to the same data shows how much a model draws on what respondents actually answered. Counting the differences reproduced, missed, and reversed for each attribute and type of question reveals which group conclusions hold, which average error alone does not show~\cite{jacobs2021measurement}. Among the conditions that meet a study's requirements, the one that discloses least can be preferred. For group comparisons that no tested condition supports, corrections that use a small set of human answers to the new questions may offer a route that keeps the record out of the prompt~\cite{krsteski2026}.

These checks fit the organizations that now build synthetic respondents from real people's data. Pew Research Center has tested personas built from its panelists' profiles~\cite{chapekis2026}. Market-research firms such as Toluna and Qualtrics sell synthetic panels built from survey respondents' data~\cite{toluna2025personas,qualtrics2026synthetic}, and startups such as Simile build agentic twins of real people for companies to survey~\cite{simile2026}. Providers like these can report group results by question type alongside population results. Data archives and ethics boards that set the terms on which survey records are used can ask which estimates a project needs the record for and what evidence shows it.
\section{Limitations and Future Work}
\label{sec:limitations}

Our results describe two U.S. surveys, one on economic inequality and one on national identity, and questions fielded alongside the answers the models receive. That is what lets us check synthetic estimates against people, and it is also what a researcher extending a survey to new questions cannot do. For questions asked in later waves or never fielded, a respondent's earlier answers may say less and the profile may shape more of the response, so the label-following in Section~\ref{sec:disc-profile} may be larger there than we measure. Panels that field new questions to the same respondents would allow this test.

Our groups are the categories each survey records, such as a first-mentioned ethnic origin and a single religious affiliation, which compress identities that respondents may hold in combination~\cite{andrus2022}. We also audit only differences the survey itself supports, which leaves out many comparisons involving small groups and intersections of attributes. These may be the groups for which the models' associations are thinnest, so the flattening we report may understate what they would face. Studying them requires larger or targeted samples. We score the response distribution each model gives, using one prompt format with the profile in the user turn, and the method used to elicit closed-ended answers can change simulated survey responses~\cite{ahnert2026survey}. Placing the profile in a system message may change how strongly models condition on it~\cite{neumann2025position}. Drawing one answer per respondent, as Pew's evaluation does, can turn attitudes common in a group into near-unanimity~\cite{chapekis2026}, and other elicitation methods may likewise enlarge the small differences we find~\cite{chen2026syntheticusers,cheng2023compost}. We draw one reassignment and one set of external profiles, follow one reduction path and run hosted models on the public survey only. Our split of questions into circumstances and attitudes is by wording and exploratory. A direct test would measure, for each question, how much of the human variation the profile explains and check whether the models' errors follow it. The statistical reference is a supervised benchmark for the same inputs, and language models fine-tuned on survey answers may come closer to it~\cite{suh2025finetune}. Our identifiability counts describe how many people share a respondent's values and do not model what an adversary knows. Finally, we measure errors in group estimates, not how researchers act on them or how the groups described experience their use. Which errors matter, and to whom, should be studied with those groups~\cite{whitney2024}.

\section{Conclusion}
\label{sec:conclusion}

We asked when a respondent's demographic record should be passed to a language model that answers survey questions in their place. The answer depends on the estimate. Population and subgroup estimates gain little from the record, and for the open-weight models most of that gain survives replacing it with census data or a synthetic persona, so for these estimates the record can stay with the analyst. Group comparisons need the record, yet with it the models make groups look more alike than they are, keeping differences in circumstances while flattening or reversing differences in attitudes, most sharply for religious and ethnic groups. Their group differences rest more on the demographic label in the prompt than on the individual respondent. Survey organizations, market-research firms and startups now build synthetic respondents from real people's data~\cite{chapekis2026,toluna2025personas,qualtrics2026synthetic,simile2026}, and such panels are often validated by average agreement with surveys or by reproducing known demographic patterns, neither of which alone establishes the validity of their group conclusions. Disclosing a record and trusting the group conclusions built on it are separate decisions, and each needs its own evidence.

\section*{Generative AI Usage Statement}
We used AI assistants, namely Claude Opus 5.5 and GPT-6 Astra, to assist with coding, checking grammar, wordsmithing, and shortening the text to fit the page limit. The authors designed the study, specified every analysis, checked code outputs and all reported numbers against the saved results, and take full responsibility for the content.
\section*{Ethical Considerations Statement}
This study is a secondary analysis of existing survey data and involved no contact with respondents. Respondents consented to a survey, and passing their records to a language model is a use that consent did not anticipate. Several of our conditions make this use, including placing one respondent’s record in another’s prompt in \textsc{Shuffled}, and we limited it in three ways. The ISSP data are used under the GESIS terms of use and were processed only with open-weight models on our own machines, so no ISSP record or prompt left them. Hosted models received only Pew prompts. The Pew data are Wave~54 of the American Trends Panel, released publicly with OpinionQA, so sending them to a hosted provider disclosed nothing beyond the public release, although many respondents are unique in that file (Appendix~\ref{app:identifiability}). Identifiability is estimated from ACS microdata and the public Wave~54 file and reported only in aggregate. The example prompts in the paper describe invented respondents, and we release code but no respondent-level ISSP prompts or predictions.

\section*{Adverse Impact Statement}
Our results could be cited as license for synthetic surveys. The finding that population estimates need little of the respondent’s record could be read as evidence that such estimates are sound, or that synthetic respondents can replace human ones, although the models still miss population distributions by wide margins and most group differences. The audit we outline could likewise become a checkbox. It tests questions a survey has already fielded, and passing it does not show that a model will hold up on new questions, where label-following may be stronger (Section~\ref{sec:limitations}). Recommending fewer fields could also be taken as a privacy guarantee, but four fields beside the answers in a prompt still single out nine in ten respondents in the public file (Appendix~\ref{app:identifiability}). We report results separately for each estimate and attribute, so that no single number stands for the whole.

\bibliographystyle{ACM-Reference-Format}
\bibliography{sample-base}

\appendix
\appendix
\normalsize\normalfont
\FloatBarrier

\section{Data, Prompts and Models}
\label{app:data}

This section defines the survey samples, profile fields, history selection, prompts, and model scoring.

\subsection{Surveys and Profile Fields}
\label{app:fields}

The study uses 63 Pew panel targets and 27 ISSP targets, with 15 demographic profile fields in each survey.
The Pew data are Wave 54 of the American Trends Panel, fielded in September 2019 and distributed with OpinionQA~\cite{pew2019atpw54,santurkar2023whose}.
We use 1,000 respondents drawn uniformly at random from 6,088 eligible English-language respondents with a complete recorded profile and positive survey weight.
The ISSP data are the 2023 National Identity and Citizenship module~\cite{issp2023}, with all 1,513 U.S. respondents who have a complete interview and recorded core demographics.
Pew targets are single-choice questions with verified routing, wording and response options that do not overlap with profile fields.
ISSP targets were sampled across thematic blocks, stratified by the spread of responses across options, among questions with at least 192 valid U.S. answers.
A target answer is evaluated when it is one of the substantive response options.
Table~\ref{tab:app-targets} lists the targets and the response side $S_q$ used in group contrasts.
Full response options follow the questionnaires~\cite{pew2019atpw54,issp2023}.

\begingroup
\scriptsize
\setlength{\tabcolsep}{3pt}
\renewcommand{\arraystretch}{0.9}
\begin{longtable}{@{}lll@{}}
\caption{Target questions and the response side $S_q$ counted in group contrasts. \textendash{} marks targets without a defined response side, which enter individual and population measures but not group contrasts.}\label{tab:app-targets}\\
\toprule
ID & Short wording & Counted response side ($S_q$)\\
\midrule
\endfirsthead
\multicolumn{3}{l}{\tablename~\thetable\ continued}\\
\toprule
ID & Short wording & Counted response side ($S_q$)\\
\midrule
\endhead
\midrule
\multicolumn{3}{r}{Continued on next page}\\
\endfoot
\bottomrule
\endlastfoot
\multicolumn{3}{@{}l}{\textbf{Pew panel}}\\
\midrule
\texttt{DEBTa\_W54} & Credit card debt & No, do not have this\\
\texttt{DEBTb\_W54} & Car loan & No, do not have this\\
\texttt{DEBTc\_W54} & Student loans & No, do not have this\\
\texttt{DEBTd\_W54} & Mortgage or home loan & No, do not have this\\
\texttt{DEBTe\_W54} & Medical-bill debt & No, do not have this\\
\texttt{ECON1B\_W54} & National economy in one year & Worse\\
\texttt{ECON1\_W54} & National economy now & Only fair / Poor\\
\texttt{ECON3\_a\_W54} & Economy factor: jobs & Not too much / Not at all\\
\texttt{ECON3\_b\_W54} & Economy factor: stock market & Not too much / Not at all\\
\texttt{ECON3\_c\_W54} & Economy factor: consumer prices & Not too much / Not at all\\
\texttt{ECON3\_d\_W54} & Economy factor: health-care costs & Not too much / Not at all\\
\texttt{ECON3\_e\_W54} & Economy factor: real estate & Not too much / Not at all\\
\texttt{ECON3\_f\_W54} & Economy factor: gas prices & Not too much / Not at all\\
\texttt{ECON3\_g\_W54} & Economy factor: federal deficit & Not too much / Not at all\\
\texttt{ECON3\_h\_W54} & Economy factor: wages/incomes & Not too much / Not at all\\
\texttt{ECON3\_i\_W54} & Economy factor: tax system & Not too much / Not at all\\
\texttt{FINANCEa\_W54} & Savings account & No, do not have this\\
\texttt{FINANCEb\_W54} & Retirement account & No, do not have this\\
\texttt{FINANCEc\_W54} & Other stocks, bonds, mutual funds & No, do not have this\\
\texttt{FIN\_SITCOMM\_W54} & Local residents' finances & Just meet their basic expenses / Don't even have enough to meet basic expenses\\
\texttt{FIN\_SITFUT\_W54} & Future household comfort & No\\
\texttt{FIN\_SITGROWUP\_W54} & Family finances growing up & Just met basic expenses / Didn't even have enough to meet basic expenses\\
\texttt{FIN\_SITMOST\_W54} & Most Americans' finances & Just meet their basic expenses / Don't even have enough to meet basic expenses\\
\texttt{FIN\_SIT\_W54} & Household finances & Just meet your basic expenses / Don't even have enough to meet basic expenses\\
\texttt{GOVPRIORITYa\_W54} & Priority: Affordable health care & Not too important / Should not be done\\
\texttt{GOVPRIORITYb\_W54} & Priority: Illegal immigration & Not too important / Should not be done\\
\texttt{GOVPRIORITYc\_W54} & Priority: Economic inequality & Not too important / Should not be done\\
\texttt{GOVPRIORITYd\_W54} & Priority: Climate change & Not too important / Should not be done\\
\texttt{GOVPRIORITYe\_W54} & Priority: Terrorism & Not too important / Should not be done\\
\texttt{GOVPRIORITYf\_W54} & Priority: Gun violence & Not too important / Should not be done\\
\texttt{GOVRESP\_a\_W54} & Federal role: Living standard & No, not the responsibility of the federal government to provide\\
\texttt{GOVRESP\_b\_W54} & Federal role: Housing & No, not the responsibility of the federal government to provide\\
\texttt{GOVRESP\_c\_W54} & Federal role: K--12 education & No, not the responsibility of the federal government to provide\\
\texttt{GOVRESP\_d\_W54} & Federal role: College education & No, not the responsibility of the federal government to provide\\
\texttt{GOVRESP\_e\_W54} & Federal role: Retirement income & No, not the responsibility of the federal government to provide\\
\texttt{GOVRESP\_f\_W54} & Federal role: Medical care & No, not the responsibility of the federal government to provide\\
\texttt{GOVRESP\_g\_W54} & Federal role: Health insurance & No, not the responsibility of the federal government to provide\\
\texttt{GOVRESP\_h\_W54} & Federal role: High-speed internet & No, not the responsibility of the federal government to provide\\
\texttt{INEQ3\_W54} & System change for inequality & Needs to be completely rebuilt\\
\texttt{INEQ4\_a\_W54} & Inequality responsibility: federal & Only a little / None\\
\texttt{INEQ4\_b\_W54} & Inequality responsibility: states & Only a little / None\\
\texttt{INEQ4\_c\_W54} & Inequality responsibility: firms & Only a little / None\\
\texttt{INEQ4\_d\_W54} & Inequality responsibility: religion & Only a little / None\\
\texttt{INEQ4\_e\_W54} & Inequality responsibility: wealthy & Only a little / None\\
\texttt{INEQ5\_a\_W54} & Inequality: job outsourcing & Contributes not too much / Contributes not at all\\
\texttt{INEQ5\_b\_W54} & Inequality: trade policies & Contributes not too much / Contributes not at all\\
\texttt{INEQ5\_c\_W54} & Inequality: job automation & Contributes not too much / Contributes not at all\\
\texttt{INEQ5\_d\_W54} & Inequality: corporate regulation & Contributes not too much / Contributes not at all\\
\texttt{INEQ5\_e\_W54} & Inequality: too little corporate regulation & Contributes not too much / Contributes not at all\\
\texttt{INEQ5\_f\_W54} & Inequality: differences in effort & Contributes not too much / Contributes not at all\\
\texttt{INEQ5\_g\_W54} & Inequality: tax system & Contributes not too much / Contributes not at all\\
\texttt{INEQ5\_h\_W54} & Inequality: life choices & Contributes not too much / Contributes not at all\\
\texttt{INEQ5\_i\_W54} & Inequality: initial opportunities & Contributes not too much / Contributes not at all\\
\texttt{INEQ5\_j\_W54} & Inequality: education system & Contributes not too much / Contributes not at all\\
\texttt{INEQ5\_k\_W54} & Inequality: racial/ethnic discrimination & Contributes not too much / Contributes not at all\\
\texttt{INEQ5\_l\_W54} & Inequality: illegal immigration & Contributes not too much / Contributes not at all\\
\texttt{INEQ5\_m\_W54} & Inequality: legal immigration & Contributes not too much / Contributes not at all\\
\texttt{INEQ6\_W54} & Family finances, success nationally & Not too much / Not at all\\
\texttt{INEQ7\_W54} & Family finances, success locally & Not too much / Not at all\\
\texttt{WORRY2a\_W54} & Worry about debt & Rarely / Never\\
\texttt{WORRY2c\_W54} & Worry about bills & Rarely / Never\\
\texttt{WORRY2d\_W54} & Worry about health-care costs & Rarely / Never\\
\texttt{WORRY2e\_W54} & Worry about retirement saving & Rarely / Never\\
\midrule
\multicolumn{3}{@{}l}{\textbf{ISSP}}\\
\midrule
\texttt{v33} \texttt{PAYTAXES} & Citizenship: not evading taxes & 5--7 (7 = Very important)\\
\texttt{v37} \texttt{CONSCRP} & Citizenship: military service & 5--7 (7 = Very important)\\
\texttt{v39} \texttt{DEM10PST\_} & Democracy ten years ago & 6--10 (10 = Very Well)\\
\texttt{v43} \texttt{POLOPTS} & More public decision participation & 5--7 (7 = Very important)\\
\texttt{v28} \texttt{IMMJOBS1} & Immigrants take American jobs & Agree strongly / Agree\\
\texttt{v32} \texttt{IMMASSIM} & Immigrants and cultural adaptation & --\\
\texttt{v27} \texttt{IMMAMECO} & Immigrants benefit the economy & Agree strongly / Agree\\
\texttt{v20} \texttt{IMPORTS} & Limit imports to protect economy & Agree strongly / Agree\\
\texttt{v25} \texttt{INTLINCS} & International firms harm business & Agree strongly / Agree\\
\texttt{v24} \texttt{AMTV} & Prefer American films on television & Agree strongly / Agree\\
\texttt{v59} \texttt{CANTRUST} & Trust in other people & People can almost always be trusted / People can usually be trusted\\
\texttt{v58} \texttt{BEFAIR} & Fairness of other people & Try to be fair most of the time / Try to be fair almost all of the time\\
\texttt{v6} \texttt{AMANCSTR} & American identity: ancestry & Very important / Fairly important\\
\texttt{v4} \texttt{AMGOVT} & American identity: institutions & Very important / Fairly important\\
\texttt{v8} \texttt{AMSHAMED} & Feeling ashamed of America & Agree strongly / Agree\\
\texttt{v16} \texttt{PROUDECO} & Pride: economic achievements & Very proud / Somewhat proud\\
\texttt{v18} \texttt{PROUDART} & Pride: arts and literature & Very proud / Somewhat proud\\
\texttt{v17} \texttt{PROUDSPT} & Pride: sports & Very proud / Somewhat proud\\
\texttt{v47} \texttt{POLEFF11} & People like me have no say & Strongly agree / Agree\\
\texttt{v48} \texttt{POLEFF18} & Government cares little about me & Strongly agree / Agree\\
\texttt{v57} \texttt{BIASNEWS} & Political news is biased & Strongly agree / Agree\\
\texttt{v56} \texttt{POLNEWS} & Frequency of political news use & --\\
\texttt{v46} \texttt{LEFTRGHT} & Political left--right placement & --\\
\texttt{v55} \texttt{EXPERTS} & Experts over ordinary experience & Strongly agree / Agree\\
\texttt{v52} \texttt{POLTALK} & Politicians talk more than act & Strongly agree / Agree\\
\texttt{v54} \texttt{BIZMONEY} & Business profits at others' expense & Strongly agree / Agree\\
\texttt{v50} \texttt{CORRUPTN} & Corruption in public service & A lot of people are involved / Almost everyone is involved\\
\end{longtable}
\endgroup

Table~\ref{tab:fields} lists the Pew panel profile fields and the fields kept in each reduced profile.
The ISSP profile contains age, sex, education, work status, religious affiliation group, religious service attendance, place of living, household income, ethnic origin (first and second mention), marital status, region, citizenship, household size and occupation major group. Reduced and external profiles are defined for the Pew panel only.

\begin{table}[h!]
\centering
\footnotesize
\caption{Pew panel profile fields included in each profile. \textsc{Shuffled-10} and \textsc{PUMS-10} use the fields of \textsc{Own-10}, and \textsc{Nemotron-4} uses those of \textsc{Own-4}.}
\label{tab:fields}
\begin{tabular}{@{}lcccc@{}}
\toprule
Field & \textsc{Own} & \textsc{Own-10} & \textsc{Own-7} & \textsc{Own-4} \\
\midrule
Census region & $\checkmark$ & $\checkmark$ & $\checkmark$ & $\checkmark$ \\
Sex & $\checkmark$ & $\checkmark$ & $\checkmark$ & $\checkmark$ \\
Age group & $\checkmark$ & $\checkmark$ & $\checkmark$ & $\checkmark$ \\
Education & $\checkmark$ & $\checkmark$ & $\checkmark$ & $\checkmark$ \\
Race or ethnicity & $\checkmark$ & $\checkmark$ & $\checkmark$ &  \\
Citizenship & $\checkmark$ & $\checkmark$ & $\checkmark$ &  \\
Place of birth & $\checkmark$ & $\checkmark$ & $\checkmark$ &  \\
Marital status & $\checkmark$ & $\checkmark$ &  &  \\
Employment status & $\checkmark$ & $\checkmark$ &  &  \\
Household income & $\checkmark$ & $\checkmark$ &  &  \\
Metropolitan residence & $\checkmark$ &  &  &  \\
Parent or guardian & $\checkmark$ &  &  &  \\
Religious affiliation & $\checkmark$ &  &  &  \\
Religious service attendance & $\checkmark$ &  &  &  \\
Born-again identification & $\checkmark$ &  &  &  \\
\midrule
Number of fields & 15 & 10 & 7 & 4 \\
\bottomrule
\end{tabular}
\end{table}

\subsection{History Selection}

\label{app:history-rule}

For each target question, the history consists of eight of the respondent's answers to other questions in the same survey,
chosen by a fixed rule that depends on the order of the questions and on which ones the respondent answered, never on the
answers themselves. In the Pew panel, we step through the questions asked of every respondent in a fixed circular order starting after the target, skip
questions from the target's question family and questions the respondent did not answer substantively, and take one question
per remaining family before filling the remaining places in the same order. The ISSP uses the same circular scan over its 27
questions without the family step, and in its one paired item (\texttt{v47}/\texttt{v48}) excludes a target's companion and includes at most one of the two. Every
respondent--target cell in the study has a full history of eight answers.
Every condition uses the same history for a given respondent and target, so differences between conditions come only from the
demographic input. Appendix~\ref{app:history} varies the number of history items.

\subsection{Prompts}

\label{app:prompts}

All conditions use a template containing the instruction, profile, earlier answers, target question, response options and answer format.
Figures~\ref{fig:app-prompt} and~\ref{fig:app-profile-blocks} show the prompt and profile variants for an invented respondent.
The survey country appears in every profile block except \textsc{Nemotron-All}, which gives a U.S. state and city.
ISSP prompts name the International Social Survey Programme and use its field labels and question wording.
Chat templates follow each model's own format.

\par
\providecommand{\promptlabel}[1]{{\normalfont\sffamily\scriptsize\color{black!55}#1}}
\providecommand{\promptgap}{\par\vspace{0.45em}}
\providecommand{\promptstyle}{\ttfamily\footnotesize\frenchspacing\raggedright\setlength{\parskip}{0pt}\linespread{0.97}\selectfont}

\par\medskip\noindent
\begin{tcolorbox}[breakable,colback=black!3,colframe=black!45,boxrule=0.4pt,arc=1pt,left=6pt,right=6pt,top=5pt,bottom=5pt]
\promptstyle
\textbf{\#\#\# Instruction}\par
\promptgap
You are participating in the Pew Research Center American Trends Panel. The personal information and previous survey answers below describe you as a survey respondent. Answer the final survey question as this respondent.\par
\promptgap
\textbf{\#\#\# Personal Information}\hfill\promptlabel{profile block}\par
\promptgap
Country: United States\par
Census region: Midwest\par
Metropolitan residence: Metropolitan\par
Sex: Female\par
Age group: 30-49\par
Education: College graduate/some postgrad\par
Race or ethnicity: Black\par
Citizenship: Yes\par
Place of birth: U.S.\par
Marital status: Married\par
Parent or guardian: No\par
Employment status: Full-time\par
Household income: \$50,000-\$75,000\par
Religious affiliation: Nothing in particular\par
Religious service attendance: Never\par
Born-again identification: No, not born-again or evangelical Christian\par
\promptgap
\textbf{\#\#\# Previous Survey Answers}\hfill\promptlabel{history (8 answers)}\par
\promptgap
\textbf{\#\#\#\# Previous answer 1}\par
\promptgap
How much, if at all, do you think the availability of jobs is contributing to your opinion about how the economy is doing?\par
\promptgap
Response options: 1 A great deal; 2 A fair amount; 3 Not too much; 4 Not at all\par
Selected option code: 2\par
\promptgap
\textbf{\#\#\#\# Previous answer 2}\par
\promptgap
Do you have a savings account?\par
\promptgap
Response options: 1 Yes, have this; 2 No, do not have this\par
Selected option code: 1\par
\promptgap
{\normalfont\itshape [Previous answers 3--7 follow the same format.]}\par
\promptgap
\textbf{\#\#\#\# Previous answer 8}\par
\promptgap
Do you think an adequate standard of living is something the federal government has a responsibility to provide for all Americans?\par
\promptgap
Response options: 1 Yes, a responsibility of the federal government to provide for all Americans; 2 No, not the responsibility of the federal government to provide\par
Selected option code: 1\par
\promptgap
\textbf{\#\#\# Question}\hfill\promptlabel{target}\par
\promptgap
Thinking about the nation\textquotesingle{}s economy, how would you rate economic conditions in this country today?\par
\promptgap
\textbf{\#\#\# Response Options}\par
\promptgap
1 Excellent; 2 Good; 3 Only fair; 4 Poor\par
\promptgap
\textbf{\#\#\# Response}\par
\promptgap
Return exactly one JSON object and no other text:\par
\{\textquotedbl{}option\textquotedbl{}: \textquotedbl{}<selected option>\textquotedbl{}\}\par
\end{tcolorbox}
\captionof{figure}{\textsc{Own} prompt excerpt for an invented Pew respondent, sent to Mistral 7B as one user message without a system message.
Labels on the right are for display.
Other conditions replace the profile block (Figure~\ref{fig:app-profile-blocks}).}
\label{fig:app-prompt}
\par\medskip

\par
\par\medskip\noindent\begin{minipage}{\linewidth}
\begin{minipage}[t]{0.48\linewidth}
\vspace{0pt}
\begin{tcolorbox}[colback=black!3,colframe=black!45,boxrule=0.4pt,arc=1pt,left=5pt,right=5pt,top=4pt,bottom=4pt,title={\small \textsc{History}},coltitle=black,colbacktitle=black!10,fonttitle=\sffamily]
\promptstyle
\textbf{\#\#\# Personal Information}\par
\promptgap
Country: United States\par
\end{tcolorbox}
\vspace{4pt}
\begin{tcolorbox}[colback=black!3,colframe=black!45,boxrule=0.4pt,arc=1pt,left=5pt,right=5pt,top=4pt,bottom=4pt,title={\small \textsc{Own-4}},coltitle=black,colbacktitle=black!10,fonttitle=\sffamily]
\promptstyle
\textbf{\#\#\# Personal Information}\par
\promptgap
Country: United States\par
Census region: Midwest\par
Sex: Female\par
Age group: 30-49\par
Education: College graduate/some postgrad\par
\end{tcolorbox}
\end{minipage}\hfill
\begin{minipage}[t]{0.48\linewidth}
\vspace{0pt}
\begin{tcolorbox}[colback=black!3,colframe=black!45,boxrule=0.4pt,arc=1pt,left=5pt,right=5pt,top=4pt,bottom=4pt,title={\small \textsc{Nemotron-All}},coltitle=black,colbacktitle=black!10,fonttitle=\sffamily]
\promptstyle
\textbf{\#\#\# Personal Information}\par
\promptgap
Sex: Male\par
Age: 67\par
Marital status: Widowed\par
Education level: Bachelor\textquotesingle{}s degree\par
Bachelor\textquotesingle{}s field: Arts and humanities\par
Occupation: Not in workforce\par
City: San Diego\par
State: CA\par
ZIP code: 92101\par
\end{tcolorbox}

\end{minipage}
\captionof{figure}{Profile blocks for the same invented respondent under \textsc{History}, \textsc{Own-4} and \textsc{Nemotron-All}.
\textsc{Own-10} and \textsc{Own-7} retain the fields marked in Table~\ref{tab:fields}.
\textsc{Shuffled}, \textsc{PUMS-10} and \textsc{Nemotron-4} use the \textsc{Own}, \textsc{Own-10} and \textsc{Own-4} formats with values from another respondent, a census record or a persona.}
\label{fig:app-profile-blocks}
\end{minipage}\par\medskip

\FloatBarrier

\subsection{Models and Scoring}

\label{app:models}

Table~\ref{tab:models} lists the model checkpoints. Open-weight models were run unquantized on NVIDIA B200 GPUs.

For an open-weight model, let $a_k$ be the complete JSON answer the prompt requests, containing the code of option $k$, and $x$ be the prompt. We score
\begin{equation*}
  s_k = \sum_{t=1}^{|a_k|} \log p_\theta\!\left(a_{k,t} \mid x, a_{k,<t}\right), \qquad
  \hat p(k \mid x) = \frac{\exp(s_k)}{\sum_{j} \exp(s_j)},
\end{equation*}
over the valid options $j$ of the question. Each $a_k$ is the complete JSON answer, closing characters included, so $s_k$ is the 
log-probability of a complete string and needs no length normalization.

GPT-4.1 and GPT-6.1 Terra label response options with letters and request the letter alone.
We normalize first-token probabilities over option letters returned among the most probable tokens, 20 for GPT-4.1 and 5 for Terra.
Letters not returned receive zero.
For Claude Haiku 5.5, we request a probability $v_k$ for each option and set $\hat p(k\mid x)=v_k/\sum_jv_j$.
We compare Claude with the other models on the direction of effects.

\begin{table}[h]
\centering
\footnotesize
\caption{Model checkpoints. Hosted models are identified by the API version used.}
\label{tab:models}
\begin{tabular}{@{}ll@{}}
\toprule
Open-weight & Hosted \\
\midrule
\texttt{google/gemma-4-12B-it} & \texttt{gpt-4.1-2025-04-14} \\
\texttt{google/gemma-4-31B-it} & \texttt{gpt-6.1-terra} \\
\texttt{allenai/Olmo-3-7B-Instruct} & \texttt{claude-haiku-5-5} \\
\texttt{allenai/Olmo-3.1-32B-Instruct} & \\
\texttt{Qwen/Qwen2.5-7B-Instruct} & \\
\texttt{Qwen/Qwen2.5-72B-Instruct} & \\
\texttt{Qwen/Qwen3.5-9B} & \\
\texttt{Qwen/Qwen3.5-27B} & \\
\texttt{meta-llama/Llama-3.1-8B-Instruct} & \\
\texttt{meta-llama/Llama-3.1-70B-Instruct} & \\
\texttt{mistralai/Mistral-7B-Instruct-v0.3} & \\
\bottomrule
\end{tabular}
\end{table}

\subsection{Reassigned and External Profiles}
\label{app:external}

External and reassigned profiles supply field values independently of the respondent's demographics and history.
\textsc{Shuffled} permutes whole recorded profiles within each survey, including missing fields.
The Pew permutation is unrestricted, so two respondents (0.2\%) receive their own profile.
The ISSP permutation permits no self-donors.
\textsc{Shuffled-10} uses the same permutation.
Each respondent receives one external profile of each kind, held fixed across all targets and models. \textsc{PUMS-10} draws, with replacement and in proportion to person weight, the record of an adult living in a household in  50 U.S. states or the District of Columbia from the 2019 ACS, with household attributes taken from the matching housing record. \textsc{Nemotron-4} and \textsc{Nemotron-All} share one persona per respondent, drawn uniformly with replacement from the adult personas in 50 U.S. states or the District of Columbia, with no weighting or demographic matching. \textsc{Nemotron-All} supplies nine structured fields, including state, exact age, occupation, and city.
ACS and Nemotron values are mapped to the matching survey categories, with state mapped to census region.

\FloatBarrier
\section{Measures and Inference}
\label{app:measures}

\textbf{Group comparisons.} For each target with a defined response side, we compare every pair of categories within an attribute and target with a defined response side. Each group must have at least 30 valid target responses and Kish effective sample size of at least 20 under the original survey weights. Missing and refused attribute values are excluded. ISSP age groups are 18--29, 30--44, 45--59 and 60 or older. The primary analysis uses 1,281 human-supported Pew pairs on prompted attributes and 1,268 ISSP pairs. Each such difference stands for a conclusion the survey supports, and we ask whether the synthetic survey supports the same conclusion.

\phantomsection
\label{app:inference}
\textbf{Reference model.} For target $q$ and condition $Z$, let $x^{Z}_{iq}$ be the one-hot encoding of respondent $i$'s history and the profile fields supplied under $Z$. The reference predicts
\[
\hat p^{\,\mathrm{ref}}_{iqk} = \frac{\exp(\beta_{qk}^{\top} x^{Z}_{iq})}{\sum_{k'} \exp(\beta_{qk'}^{\top} x^{Z}_{iq})},
\]
with coefficients fit on the four folds that exclude respondent $i$ by minimizing survey-weighted log loss plus an L2 penalty. No respondent's answer to $q$ is used to predict it. The penalty is chosen from three values separately for each target, condition, and outer fold by three-fold validation log loss, with ties favoring the stronger penalty. Evaluation uses equal respondent weights.

\FloatBarrier
\newpage

\section{Additional Results}

This section reports history-length, profile, hosted-model, group-conclusion, question-type, and identifiability results.

\subsection{History Length}

\label{app:history}

\begin{figure*}[h]
\centering
\includegraphics[width=0.9\textwidth]{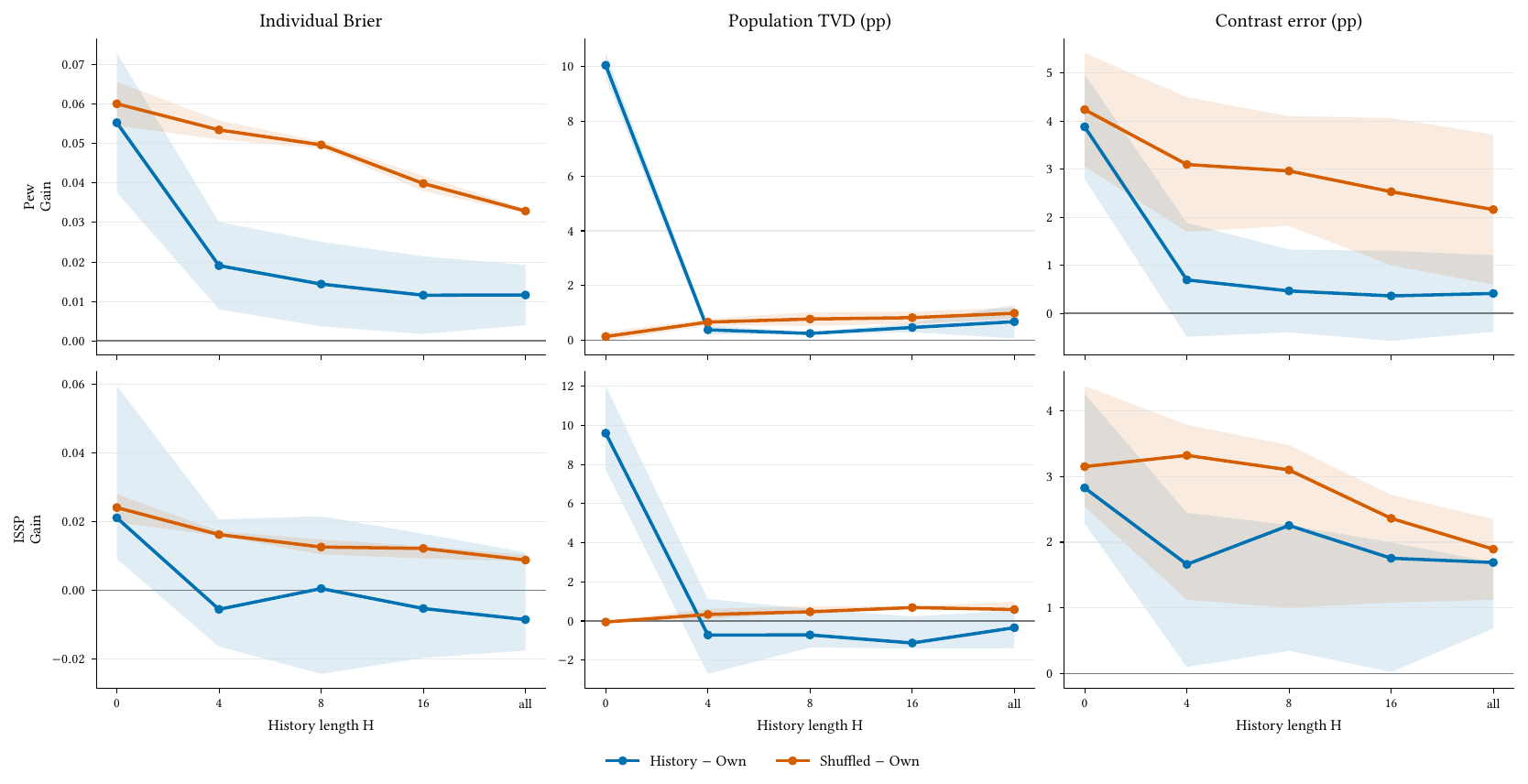}
\caption{Profile gains (\textsc{History} to \textsc{Own}) and linkage gains (\textsc{Shuffled} to \textsc{Own}) by history length.
Lines are medians over the two Pew and three ISSP models for each length, with model ranges shaded.
Group comparisons use 1,281 Pew and 1,268 ISSP human-supported pairs on prompted attributes.
Positive gains indicate lower error.}
\label{fig:app-history}
\end{figure*}

Figure~\ref{fig:app-history} reports profile and linkage gains as history length varies.
\textsc{History}, \textsc{Own} and \textsc{Shuffled} use 0, 4, 8, 16 or all available answers, with each shorter history a prefix of the longer one.
Llama~3.1 70B and Qwen2.5 72B have every length in both surveys, and Gemma~4 12B also has every length in the ISSP. Seven models were also run with no history, and Table~\ref{tab:app-h0} reports their group comparisons and population gains at zero and eight answers.

\begin{table*}[!htbp]
\centering
\footnotesize
\setlength{\tabcolsep}{4pt}
\caption{Reproduced differences and population gains with no history and eight answers on the same seven-model panel.}
\label{tab:app-h0}
\begin{tabular}{@{}lrrrrrrrrr@{}}
\toprule
 & \multicolumn{3}{c}{$H=0$: reproduced (\%)}
 & \multicolumn{3}{c}{$H=8$: reproduced (\%)}
 & \multicolumn{2}{c}{$H=0$: TVD gain (pp)}
 & \shortstack{Own overlap\\(\%)} \\
\cmidrule(lr){2-4}\cmidrule(lr){5-7}\cmidrule(lr){8-9}
Survey (pairs) & History & Own & Shuffled & History & Own & Shuffled
 & Own & Shuffled & $H=8 \cap H=0$ \\
\midrule
Pew (1,281) & 0.0 & 40.4 & 1.2 & 16.5 & 41.8 & 9.0 & 10.24 & 10.21 & 76.3 \\
ISSP (1,268) & 0.0 & 33.5 & 0.6 & 15.8 & 34.9 & 11.8 & 7.93 & 8.08 & 64.3 \\
Both surveys & \multicolumn{8}{c}{---} & 71.0 \\
\bottomrule
\end{tabular}
\par\smallskip
\begin{minipage}{\textwidth}
\footnotesize
Shares and gains are medians of model-specific statistics.
Overlap pools model-pair counts and divides the intersection by Own reproductions at $H=8$. The combined row also pools surveys.
All comparisons group predictions by respondents' own attributes.
The seven models are Gemma~4 12B, OLMo~3 7B, Qwen2.5 7B and 72B, Qwen3.5 9B, and Llama~3.1 8B and 70B.
\end{minipage}
\end{table*}

\FloatBarrier
\newpage

\subsection{Unlinked and Reduced Profiles}
\label{app:alternatives}

Table~\ref{tab:app-alternatives} reports gains over \textsc{History} for unlinked and reduced Pew profiles.

\begin{table*}[!htbp]
\centering
\small
\setlength{\tabcolsep}{3.5pt}
\caption{Gains over \textsc{History} for Pew profiles.
Model cells give medians across eleven models with improves/worsens/unresolved counts from basic 95\% intervals.
Ref. denotes the reference fit with the same fields.
Positive gains indicate lower error.
Brier gains are multiplied by 100 and other gains are in pp.
Reproduced shares and contrast errors use 1,281 human-supported pairs on prompted attributes.}
\label{tab:app-alternatives}
\begin{tabular}{@{}l rr rr rr rr rr@{}}
\toprule
& \multicolumn{2}{c}{Individual Brier ($\times100$)} & \multicolumn{2}{c}{Population TVD} & \multicolumn{2}{c}{Subgroup TVD} & \multicolumn{2}{c}{Contrast error} & \multicolumn{2}{c}{Reproduced share} \\
\cmidrule(lr){2-3}\cmidrule(lr){4-5}\cmidrule(lr){6-7}\cmidrule(lr){8-9}\cmidrule(l){10-11}
Condition & Models & Ref. & Models & Ref. & Models & Ref. & Models & Ref. & Models & Ref. \\
\midrule
\textsc{History} & --- & --- & --- & --- & --- & --- & --- & --- & 22.6\% & 41.8\% \\
\textsc{Own} & 2.56 (8/1/2) & 1.36 & 0.74 (6/3/2) & 0.42 & 0.37 (6/3/2) & 1.41 & 1.13 (7/1/3) & 6.66 & 41.8\% & 82.2\% \\
\midrule
\multicolumn{11}{@{}l}{\emph{Reduced own profile}} \\
\textsc{Own-10} & 2.50 (9/2/0) & 1.23 & 1.07 (6/2/3) & 0.39 & 0.66 (6/3/2) & 1.10 & 0.43 (6/4/1) & 5.32 & 40.0\% & 72.8\% \\
\textsc{Own-7} & 1.77 (8/2/1) & 1.03 & 0.43 (6/3/2) & 0.41 & 0.24 (5/4/2) & 0.96 & 0.53 (7/1/3) & 4.03 & 34.5\% & 68.1\% \\
\textsc{Own-4} & 1.31 (7/4/0) & 0.83 & 0.75 (6/4/1) & 0.50 & 0.29 (5/4/2) & 0.82 & 0.32 (6/1/4) & 3.57 & 31.3\% & 65.8\% \\
\midrule
\multicolumn{11}{@{}l}{\emph{Profiles not linked to the respondent}} \\
\textsc{Shuffled-10} & -1.95 (1/8/2) & -0.44 & 0.38 (5/4/2) & -0.01 & 0.28 (5/4/2) & -0.04 & -0.90 (0/11/0) & -0.24 & 10.7\% & 40.0\% \\
\textsc{PUMS-10} & -1.44 (2/8/1) & -0.50 & 0.81 (6/3/2) & -0.01 & 0.71 (6/4/1) & -0.02 & -0.64 (0/7/4) & -0.19 & 12.9\% & 40.3\% \\
\textsc{Nemotron-4} & -1.49 (3/7/1) & -0.26 & 0.77 (6/5/0) & -0.02 & 0.76 (6/5/0) & -0.04 & -0.30 (0/6/5) & -0.14 & 18.0\% & 40.4\% \\
\textsc{Nemotron-All} & -1.16 (3/8/0) & -0.58 & -0.53 (4/6/1) & 0.04 & -0.48 (5/6/0) & -0.04 & -0.49 (0/6/5) & -0.27 & 15.5\% & 39.9\% \\
\bottomrule
\end{tabular}
\end{table*}

\FloatBarrier

\subsection{Hosted Models}

\label{app:hosted}

Table~\ref{tab:app-hosted} reports hosted-model gains on the Pew panel under the scoring methods in Appendix~\ref{app:models}.

\begin{table*}[!htbp]
\centering
\footnotesize
\setlength{\tabcolsep}{4pt}
\caption{Hosted-model gains on the Pew panel.
Positive values indicate lower error.
Brier gains are multiplied by 100 and other gains are in pp.
Contrast errors use 1,281 human-supported pairs on prompted attributes.
Basic 95\% bootstrap verdicts are $+$ for improvement, $-$ for worsening and $\circ$ for unresolved.}
\label{tab:app-hosted}
\begin{tabular}{@{}ll rrrr@{}}
\toprule
Model & Comparison & Individual Brier & Population TVD & Subgroup TVD & Contrast error \\
\midrule
GPT-4.1 & \textsc{Own} vs.\ \textsc{History} & 5.38 $+$ & 1.22 $+$ & 0.97 $+$ & 1.24 $+$ \\
 & \textsc{Shuffled} vs.\ \textsc{History} & $-$0.91 $-$ & 0.59 $\circ$ & 0.73 $+$ & $-$1.39 $-$ \\
 & \textsc{PUMS-10} vs.\ \textsc{History} & $-$0.82 $-$ & $-$0.77 $-$ & $-$0.61 $-$ & $-$0.79 $-$ \\
 & \textsc{Own-10} vs.\ \textsc{History} & 3.32 $+$ & $-$0.69 $-$ & $-$1.03 $-$ & 0.23 $\circ$ \\
 & \textsc{Own-4} vs.\ \textsc{History} & 2.42 $+$ & $-$0.42 $-$ & $-$0.40 $-$ & 0.55 $+$ \\
 & \textsc{PUMS-10} vs.\ \textsc{Own-10} & $-$4.14 $-$ & $-$0.08 $\circ$ & 0.42 $+$ & $-$1.01 $-$ \\
\midrule
GPT-6.1 Terra & \textsc{Own} vs.\ \textsc{History} & 4.48 $+$ & 2.51 $+$ & 1.98 $+$ & 1.28 $+$ \\
 & \textsc{Shuffled} vs.\ \textsc{History} & $-$3.55 $-$ & 0.82 $+$ & 0.66 $+$ & $-$2.23 $-$ \\
 & \textsc{PUMS-10} vs.\ \textsc{History} & $-$2.54 $-$ & 0.41 $\circ$ & 0.34 $\circ$ & $-$1.81 $-$ \\
 & \textsc{Own-10} vs.\ \textsc{History} & 3.45 $+$ & 2.13 $+$ & 1.67 $+$ & 0.72 $+$ \\
 & \textsc{Own-4} vs.\ \textsc{History} & 2.53 $+$ & 2.40 $+$ & 2.03 $+$ & 1.02 $+$ \\
 & \textsc{PUMS-10} vs.\ \textsc{Own-10} & $-$5.99 $-$ & $-$1.73 $-$ & $-$1.33 $-$ & $-$2.53 $-$ \\
\midrule
Claude Haiku 5.5 & \textsc{Own} vs.\ \textsc{History} & 1.44 $+$ & 0.40 $+$ & 0.69 $+$ & 4.10 $+$ \\
 & \textsc{Shuffled} vs.\ \textsc{History} & $-$0.93 $-$ & 0.39 $+$ & 0.27 $+$ & $-$0.56 $-$ \\
 & \textsc{PUMS-10} vs.\ \textsc{History} & $-$0.90 $-$ & 0.53 $+$ & 0.43 $+$ & $-$0.24 $\circ$ \\
 & \textsc{Own-10} vs.\ \textsc{History} & 1.34 $+$ & 0.51 $+$ & 0.71 $+$ & 4.15 $+$ \\
 & \textsc{Own-4} vs.\ \textsc{History} & 0.71 $+$ & 0.62 $+$ & 0.63 $+$ & 2.33 $+$ \\
 & \textsc{PUMS-10} vs.\ \textsc{Own-10} & $-$2.24 $-$ & 0.02 $\circ$ & $-$0.28 $-$ & $-$4.40 $-$ \\
\bottomrule
\end{tabular}
\end{table*}

\FloatBarrier
\newpage
\subsection{Group Conclusions by Model}
\label{app:size}

\begin{figure*}[t]
\centering
\includegraphics[width=0.9\textwidth]{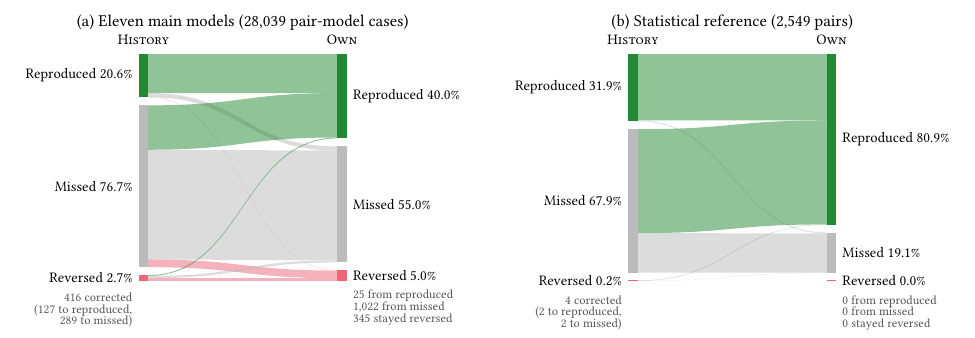}
\caption{With the profile, the models resolve nearly twice as many group differences, and about one resolved difference in nine points the wrong way under both conditions. Most differences remain missed. The reference reproduces four in five and reverses none. Each human-supported difference is traced from \textsc{History} to \textsc{Own}, pooled over both surveys: (a) the eleven main models, pooled over models, and (b) the statistical reference.}
\label{fig:transitions}
\end{figure*}

Figure~\ref{fig:transitions} traces human-supported group differences from \textsc{History} to \textsc{Own}, pooled over both surveys, for the eleven main models and the statistical reference. Table~\ref{tab:app-size} reports the results for each model and the reference under \textsc{Own}: the shares reproduced and missed, the median signed ratio, and the numbers of newly reversed and corrected differences relative to \textsc{History}. Within each of the five families run at two sizes, the larger model reproduces 6 to 18 percentage points more of the Pew differences and creates fewer new reversals pooled over both surveys. Four of the five larger open-weight models still have more new reversals than corrections pooled over both surveys. The two GPT models have fewer new reversals than corrections in the Pew panel.

\begin{table}[h!]
\centering
\scriptsize
\setlength{\tabcolsep}{2pt}
\caption{Group conclusions under \textsc{Own} on prompted attributes, by model. Reprod.\ and Missed are shares of human-supported pairs. Median $r$ is the signed synthetic-to-human contrast ratio over all human-supported pairs. New counts newly reversed pairs from \textsc{History} to \textsc{Own}, Corr.\ counts pairs reversed under History that are not reversed under \textsc{Own}, and Net is New minus Corr.\ Hosted models have Pew values only. $\dagger$ Claude Haiku 5.5 is direction-only, as described in Appendix~\ref{app:models}.}
\label{tab:app-size}
\begin{tabular}{@{}lrrrrrrrrrrrr@{}}
\toprule
 & \multicolumn{6}{c}{Pew panel (1,281 prompted pairs), \textsc{Own}} & \multicolumn{6}{c}{ISSP (1,268 prompted pairs), \textsc{Own}} \\
\cmidrule(lr){2-7}\cmidrule(l){8-13}
Model & Reprod.\ (\%) & Missed (\%) & Median $r$ & New & Corr. & Net & Reprod.\ (\%) & Missed (\%) & Median $r$ & New & Corr. & Net \\
\midrule
Gemma~4 12B & 53.0 & 44.7 & 0.596 & 21 & 23 & -2 & 45.0 & 54.3 & 0.380 & 6 & 21 & -15 \\
Gemma~4 31B & 60.2 & 38.0 & 0.853 & 13 & 30 & -17 & 52.4 & 46.6 & 0.635 & 12 & 10 & +2 \\
\midrule
OLMo~3 7B & 29.7 & 45.1 & 0.017 & 254 & 35 & +219 & 15.1 & 75.6 & 0.012 & 82 & 29 & +53 \\
OLMo~3.1 32B & 37.5 & 57.7 & 0.148 & 44 & 18 & +26 & 38.2 & 58.4 & 0.203 & 27 & 23 & +4 \\
\midrule
Qwen2.5 7B & 32.6 & 65.2 & 0.014 & 25 & 20 & +5 & 22.6 & 73.3 & 0.050 & 49 & 17 & +32 \\
Qwen2.5 72B & 41.8 & 54.3 & 0.212 & 44 & 5 & +39 & 43.9 & 53.6 & 0.361 & 17 & 6 & +11 \\
\midrule
Qwen3.5 9B & 40.8 & 56.3 & 0.184 & 35 & 10 & +25 & 34.9 & 56.9 & 0.211 & 75 & 8 & +67 \\
Qwen3.5 27B & 59.2 & 38.6 & 0.640 & 14 & 25 & -11 & 54.3 & 41.1 & 0.485 & 41 & 3 & +38 \\
\midrule
Llama~3.1 8B & 43.2 & 46.7 & 0.267 & 123 & 19 & +104 & 18.3 & 74.4 & 0.057 & 58 & 34 & +24 \\
Llama~3.1 70B & 49.6 & 46.9 & 0.327 & 38 & 12 & +26 & 48.4 & 48.5 & 0.320 & 24 & 3 & +21 \\
\midrule
Mistral 7B & 30.2 & 66.4 & 0.043 & 17 & 44 & -27 & 29.2 & 68.0 & 0.154 & 28 & 21 & +7 \\
\midrule
GPT-4.1 & 51.8 & 46.4 & 0.525 & 10 & 21 & -11 & --- & --- & --- & --- & --- & --- \\
GPT-6.1 Terra & 58.5 & 40.6 & 0.638 & 8 & 16 & -8 & --- & --- & --- & --- & --- & --- \\
Claude Haiku 5.5$^\dagger$ & 36.2 & 63.0 & 0.208 & 10 & 1 & +9 & --- & --- & --- & --- & --- & --- \\
\midrule
Reference & 82.2 & 17.8 & 0.737 & 0 & 3 & -3 & 79.5 & 20.5 & 0.554 & 0 & 1 & -1 \\
\bottomrule
\end{tabular}
\end{table}

\FloatBarrier
\newpage
\subsection{Circumstances and Attitudes}

\label{app:circumstances}

Table~\ref{tab:circumstances} compares reproduced differences for circumstance and attitude items.
The 63 Pew targets were classified by wording into 15 circumstance and 48 attitude items before computing the summaries.
The comparison uses attributes with at least 20 human-supported pairs in each class.
The hypothesis came from comparisons by race, and religious affiliation had been examined in an earlier exploratory pass, so these two rows are not independent tests.

\begin{table}[!htbp]
  \centering
  \small
  \setlength{\tabcolsep}{4pt}
  \caption{Circumstances and attitudes under \textsc{Own}. The model pooled
  column reports reproduced / synthetic-null / reversed percentages across
  eleven models. Pairs count human-supported comparisons within each item
  class. All percentages use human-supported pairs as the denominator.
  The remaining share is uncertain.}
  \label{tab:circumstances}
  \begin{tabular}{@{}llrcr@{}}
    \toprule
    Attribute & Item class & Pairs & Model pooled (\%) &
    \shortstack{Reference\\reproduced (\%)} \\
    \midrule
    Age & Circumstances & 50 & 59.27 / 10.91 / 4.55 & 96.00 \\
        & Attitudes & 92 & 22.92 / 38.74 / 13.34 & 81.52 \\
    \addlinespace
    Education & Circumstances & 103 & 84.11 / 5.12 / 2.29 & 97.09 \\
              & Attitudes & 120 & 28.26 / 36.14 / 9.02 & 75.00 \\
    \addlinespace
    Marital status & Circumstances & 73 & 70.86 / 6.35 / 3.49 & 89.04 \\
                   & Attitudes & 115 & 27.19 / 45.14 / 4.98 & 80.00 \\
    \addlinespace
    Income & Circumstances & 92 & 83.40 / 8.00 / 3.06 & 98.91 \\
           & Attitudes & 90 & 42.12 / 32.32 / 8.99 & 83.33 \\
    \addlinespace
    Religious affiliation & Circumstances & 27 & 57.91 / 8.08 / 1.68 & 77.78 \\
                          & Attitudes & 145 & 18.43 / 46.77 / 9.09 & 76.55 \\
    \addlinespace
    Religious attendance & Circumstances & 34 & 29.95 / 10.43 / 1.87 & 70.59 \\
                         & Attitudes & 96 & 19.89 / 46.69 / 5.87 & 80.21 \\
    \addlinespace
    \shortstack[l]{Race/ethnicity\\(hypothesis source)} & Circumstances & 31 &
    82.40 / 4.69 / 3.23 & 93.55 \\
                         & Attitudes & 58 & 46.39 / 36.99 / 2.35 & 86.21 \\
    \bottomrule
  \end{tabular}
\end{table}

\begin{table}[!htbp]
  \centering
  \small
  \setlength{\tabcolsep}{4pt}
  \caption{Target-item classification. The common \texttt{\_W54} suffix is
  omitted. Ranges include every intervening letter.}
  \label{tab:circumstances-items}
  \begin{tabular}{@{}p{0.17\linewidth}p{0.79\linewidth}@{}}
    \toprule
    Item class & Item identifiers \\
    \midrule
    Circumstances (15) &
    \texttt{DEBTa}--\texttt{DEBTe},
    \texttt{FINANCEa}--\texttt{FINANCEc},
    \texttt{FIN\_SIT}, \texttt{FIN\_SITFUT},
    \texttt{FIN\_SITGROWUP}, \texttt{WORRY2a},
    \texttt{WORRY2c}--\texttt{WORRY2e}. \\
    \addlinespace
    Attitudes (48) &
    \texttt{ECON1}, \texttt{ECON1B},
    \texttt{ECON3\_a}--\texttt{ECON3\_i},
    \texttt{FIN\_SITCOMM}, \texttt{FIN\_SITMOST},
    \texttt{GOVPRIORITYa}--\texttt{GOVPRIORITYf},
    \texttt{GOVRESP\_a}--\texttt{GOVRESP\_h},
    \texttt{INEQ3}, \texttt{INEQ4\_a}--\texttt{INEQ4\_e},
    \texttt{INEQ5\_a}--\texttt{INEQ5\_m},
    \texttt{INEQ6}, \texttt{INEQ7}. \\
    \bottomrule
  \end{tabular}
\end{table}

\FloatBarrier
\newpage

\subsection{Identifiability}

\label{app:identifiability}

Table~\ref{tab:identifiability} reports how many U.S. adults share each respondent's profile values and how many respondents are unique in the public Wave 54 file, with and without the answers in a prompt.

\begin{table}[htb]
\centering
\small
\setlength{\tabcolsep}{4pt}
\caption{Population frequency and Wave 54 sample uniqueness for 1,000 Pew respondents. ACS counts use 2019 person weights and the PUMS-10 mapping. Uniqueness uses the full 6,878-person Wave 54 file. History contains the eight answers in each saved \texttt{ECON1\_W54} prompt. Dashes indicate fields without ACS counterparts. Zero ACS matches reflect PUMS resolution or unmappable refusal values and do not establish uniqueness.}
\label{tab:identifiability}
\begin{tabular}{@{}lrrrr@{}}
\toprule
Profile & \shortstack{Median ACS\\count} & \shortstack{Below 1,000\\adults (\%)} & \shortstack{Unique in Wave 54\\profile (\%)} & \shortstack{Unique in Wave 54\\profile + history (\%)} \\
\midrule
History only & -- & -- & -- & 18.0 \\
Own-4 & 1,509,304 & 0.0 & 0.1 & 90.4 \\
Own-7 & 776,456 & 0.5 & 6.5 & 93.6 \\
Own-10 & 21,211 & 13.1 & 46.6 & 99.5 \\
Own (15) & -- & -- & 94.5 & 100.0 \\
\bottomrule
\end{tabular}
\end{table}

\FloatBarrier

\end{document}